 \documentclass[twocolumn,aps,prb,superscriptaddress, amsmath,amssymb,longbibliography]{revtex4-2}
 \newcommand{\figurewidth}{\columnwidth}

\usepackage{graphicx}
\usepackage{siunitx}	
\usepackage{amsmath,amssymb}
\usepackage{datetime}
\usepackage[version=4]{mhchem} 
\usepackage{booktabs}
\usepackage{hyperref}
\usepackage[noabbrev, capitalise, nameinlink]{cleveref}
\renewcommand{\autoref}[1]{\cref{#1}}
\usepackage{xcolor}
\hypersetup{
    colorlinks,
    linkcolor={red!50!black},
    citecolor={blue!50!black},
    urlcolor={blue!80!black}
}

\usepackage{todonotes}

\newcommand{\YNP}{\texorpdfstring{YbNi$_4$P$_2$}{YbNi4P2}}					\newcommand{\LNP}{\texorpdfstring{LuNi$_4$P$_2$}{LuNi4P2}}
\newcommand{\YRS}{\texorpdfstring{YbRh$_2$Si$_2$}{YbRh2Si2}}
\newcommand{\URS}{\texorpdfstring{URu$_2$Si$_2$}{YbRh2Si2}}

\newcommand{\kT}{\kilo\tesla}

\newcommand{\me}{\ensuremath{m_{\text e}}}
\newcommand{\dd}{\text{d}}	
\newcommand{\LK}{Lifshitz-Kosevich}
\newcommand{\Biii}{\ensuremath{B_3}}
\newcommand{\Biv}{\ensuremath{B_4}}
\newcommand{\Bvii}{\ensuremath{B_7}}
\newcommand{\Bviii}{\ensuremath{B_8}}
\newcommand{\Bix}{\texorpdfstring{\ensuremath{B_9}}{B9}}

\newcommand{\Fsc}{\ensuremath{F_{\circ}^{[001]}}}
\newcommand{\Fa}{\ensuremath{F_{\alpha}}}
\newcommand{\Fap}{\ensuremath{F_{\alpha}^{+}}}
\newcommand{\Fam}{\ensuremath{F_{\alpha}^{-}}}
\newcommand{\Fapm}{\ensuremath{F_{\alpha}^{\pm}}}
\newcommand{\Fb}{\ensuremath{F_{\beta}}}
\newcommand{\Fc}{\ensuremath{F_{\gamma}}}
\newcommand{\mc}{\ensuremath{m^{\star}_{\gamma}}}
\newcommand{\mab}{\ensuremath{m^{\star}_{\alpha,\beta}}}
\newcommand{\Fcp}{\ensuremath{F_{\gamma}^{+}}}
\newcommand{\Fcm}{\ensuremath{F_{\gamma}^{-}}}
\newcommand{\Fcpm}{\ensuremath{F_{\gamma}^{\pm}}}
\newcommand{\Fd}{\ensuremath{F_{\delta}}}
\newcommand{\Fdp}{\ensuremath{F_{\delta}^{+}}}
\newcommand{\Fdm}{\ensuremath{F_{\delta}^{-}}}
\newcommand{\Fdpm}{\ensuremath{F_{\delta}^{\pm}}}
\newcommand{\Fe}{\ensuremath{F_{\epsilon}}}
\newcommand{\Ff}{\ensuremath{F_{\zeta}}}
\newcommand{\Fg}{\ensuremath{F_{\eta}}}
\newcommand{\Fh}{\ensuremath{F_{\theta}}}
\newcommand{\Fi}{\ensuremath{F_{\iota}}}
\newcommand{\mstar}{\ensuremath{m^{\star}}}
\newcommand{\mDFT}{\ensuremath{m^{\text{(b)}}}}

\newcommand{\mDFTc}{\ensuremath{m^{\text{(b)}}_{\gamma}}}
\newcommand{\mDFTab}{\ensuremath{m^{\text{(b)}}_{\alpha,\beta}}}

\newcommand{\FA}{\ensuremath{F_{A}}}
\newcommand{\Bj}{\ensuremath{B_j}}
\newcommand{\TK}{\ensuremath{T_{\text K}}}
\newcommand{\TC}{\ensuremath{T_{\text C}}}
\newcommand{\gammaS}{\ensuremath{\gamma_{\text S}}}

\newcommand{\EF}{\ensuremath{E_{\text F}}}
\newcommand{\muB}{\ensuremath{\mu_{\text{B}}}}

\newcommand{\Fo}{\ensuremath{F_0}}
\newcommand{\FApm}{\ensuremath{F_{\text A}^{\pm}}}
\newcommand{\Fbp}{\ensuremath{F_{\text{bp}}}}
\newcommand{\Fbppm}{\ensuremath{F_{\text{bp}}^{\pm}}}
\newcommand{\gl}{\ensuremath{g_{\text 1}}}

\begin{document}

%
%

\preprint{ver 1.1 \today\ \currenttime}

\title{
\textit{Field-tuned quasiparticles and electronic structure in heavy-fermion \YNP}
}

\author{Will Broad}
\affiliation{HH Wills Physics Laboratory, University of Bristol, Bristol, BS8 1TL, UK}

\author{Owen Moulding}
\affiliation{HH Wills Physics Laboratory, University of Bristol, Bristol, BS8 1TL, UK}

\author{Takaki Muramatsu}
\affiliation{HH Wills Physics Laboratory, University of Bristol, Bristol, BS8 1TL, UK}

\author{Manuel Brando}
\affiliation{Max Planck Institute for Chemical Physics of Solids,
Nöthnitzer Strasse 40, D-01187 Dresden, Germany}

\author{Alix McCollam}
\affiliation{High Field Magnet Laboratory (HFML-EMFL), Radboud University, Toernooiveld 7, Nijmegen 6525 ED, Netherlands}
\affiliation{School of Physics, University College Cork, Cork, Ireland}

\author{Femke Bangman}
\affiliation{High Field Magnet Laboratory (HFML-EMFL), Radboud University, Toernooiveld 7, Nijmegen 6525 ED, Netherlands}

\author{Gertrud Zwicknagl}
\affiliation{Institute for Mathematical Physics, Technische Universität Braunschweig, 38106 Braunschweig, Germany}
\affiliation{Max Planck Institute for Chemical Physics of Solids,
Nöthnitzer Strasse 40, D-01187 Dresden, Germany}

\author{Kristin Kliemt}
\affiliation{Physikalisches Institut, Goethe-Universität Frankfurt, Max-von-Laue-Straße 1, 60438 Frankfurt am Main, Germany}

\author{Cornelius Krellner}
\affiliation{Physikalisches Institut, Goethe-Universität Frankfurt, Max-von-Laue-Straße 1, 60438 Frankfurt am Main, Germany}

\author{Sven Friedemann}
\email{Sven.Friedemann@bristol.ac.uk}
\affiliation{HH Wills Physics Laboratory, University of Bristol, Bristol, BS8 1TL, UK}

\date{\today}



\begin{abstract}

We study the Fermi surface topology and quasiparticle properties in the heavy fermion compound \YNP\ at high magnetic fields using quantum oscillation measurements. We observe a large decrease of the quasiparticle mass with increasing field and demonstrate good qualitative agreement with the single-ion Kondo model. At the putative Lifshitz transition at \SI{17}{\tesla}, we observe a sudden change of quantum oscillation frequencies suggesting an abrupt change of the electronic structure and/or quasiparticle characteristics. Our results demonstrate the ability to tune the electronic structure and provide input for theoretical models of \YNP\ and correlated electron systems in high magnetic fields. 
\end{abstract}

\maketitle

In metallic systems, the topology of the Fermi surface together with the quasiparticle mass determine thermodynamic, magnetic, and transport properties as well as instabilities towards ordered states like superconductivity and magnetism. Thus, it is of wide interest to understand and tune the electronic structure, Fermi surface topology and quasiparticle properties. External magnetic fields are a convenient parameter for tuning as it changes the Zeeman energy of the electronic states. However, for most metals, the Zeeman energy is small compared to the Fermi energy and thus changes are small or require fields that are not accessible.

By contrast, heavy-fermion compounds respond strongly to external stimuli because of the reduced Fermi energy scale $\approx\SIrange{1}{10}{\milli\electronvolt}$ of the composite quasiparticles forming from the hybridisation of local moments and conduction electrons.  Hence, heavy-fermion materials offer a fertile ground to study the Fermi surface and quasiparticle properties in the presence of many-body effects \cite{Wirth2016}. 

Electronic quasiparticles with enhanced mass and strong renormalisation of other properties are formed in heavy-fermion compounds. Composite quasiparticles emerge from the screening of local moments (typically from 4$f$ and 3$d$ elements) by conduction electrons: the resulting Kondo singlets develop a weak dispersion and become delocalised. Thus, the Kondo singlets form heavy quasiparticles which follow the characteristics of a Fermi liquid but with the effective mass and other properties strongly renormalised.

The volume enclosed by the Fermi-surface increases in heavy-fermion compounds forming the so-called ``large'' Fermi surface when the degrees of freedom of the local moments contribute to the Fermi volume. Thus, studies of the Fermi surface topology, volume, and quasiparticle properties upon tuning the strength of the Kondo effect allows to answer fundamental questions on many-body quantum states under various tuning parameters like magnetic field \cite{Friedemann2010b} and pressure\cite{Shishido2005}.

\begin{figure*}%
\includegraphics[width=.8\textwidth]{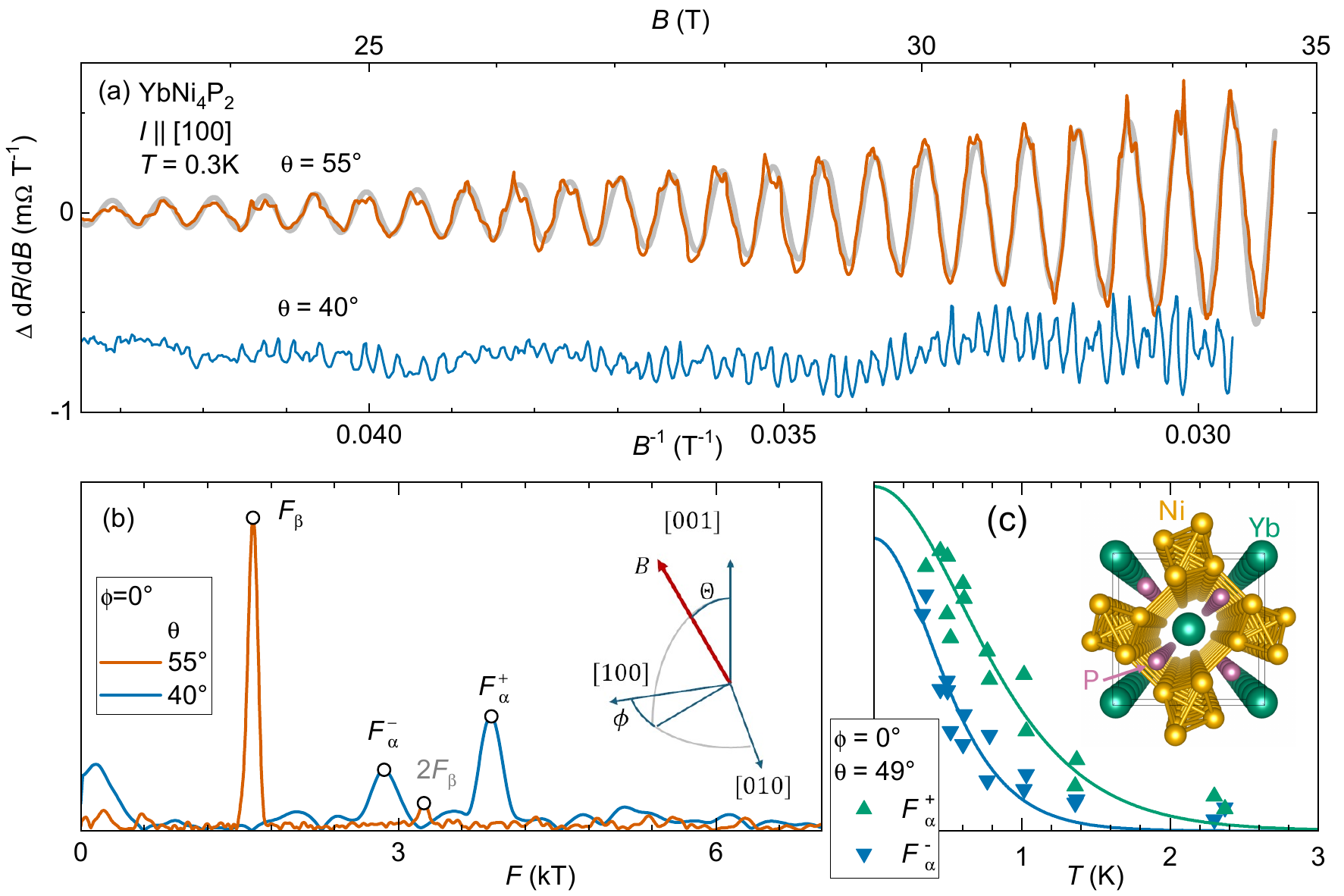}%
\caption{\textbf{Quantum oscillations in \YNP.} 
    (a) Oscillatory signal in the derivative of the magnetoresistance after removing a linear background for representative directions of the magnetic field as indicated by polar angle $\theta$\ and azimuth $\phi$.
    (b) Fast Fourier transform (FFT) for the data shown in (a). The traces at $\theta=\SI{40}{\degree}$ and  $\theta=\SI{55}{\degree}$ were analysed over the field range \SIrange{30}{35}{\tesla} and \SIrange{20}{35}{\tesla}, respectively.
				Inset shows a sketch of the field orientation relative to the  crystallographic directions.
    (c) Temperature dependence of peak amplitudes in the FFT, with Lifshitz-Kosevich fits for \Fap\ and \Fam.
		Inset shows crystal structure of \YNP\ projected along the [001] axis with the unit cell indicated by the wire frame.
    }
    \label{fig:QO}
\end{figure*}%

In large magnetic fields, the composite quasiparticles become de-renormalised because the formation of Kondo singlets is suppressed \cite{Shiina2021}.  
The combination of the suppression of the Kondo effect and the spin splitting by the Zeeman effect lead to complex behaviour of the electronic structure and quasiparticle properties in high magnetic fields. 
It remains an open question, how the Fermi surface volume, topology, quasiparticle mass, and $g$-factor evolve \cite{Hu2024}. 
Understanding composite quasiparticles in high magnetic fields is important for fundamental understanding of many-body states, the combination with topological invariants, and many prominent systems where the Fermi surface and quasiparticle mass undergo strong changes upon variation of pressure, doping, and due to symmetry-breaking states, e.g. in FeSe \cite{Terashima2016} and cuprate superconductors \cite{Sebastian2015}. Furthermore, it provides an avenue to model the high-field superconducting states in exotic superconductors like CeRh$_2$As$_2$ \cite{Khim2021} and other heavy-fermion superconductors.
Here, we study the heavy-fermion compound \YNP, which affords a direct view on the Fermi-surface and quasiparticle evolution through quantum oscillation (QO) measurements.

Whilst the suppression of the Kondo effect has been studied with various bulk probes like specific heat \cite{Pfau2013}, we use quantum oscillations as a direct probe of specific orbits on the Fermi surface and their effective masses. We compare our experimental results of the mass de-renormalisation with predictions from the single-ion Kondo model. In addition, we are able to uncover the field-dependence of the Fermi surface.

The heavy-fermion material \YNP\ has a tetragonal structure (see inset in \autoref{fig:QO}(c)) and has been synthesised recently with high purity enabling detailed quantum oscillation measurements as evident from the long mean free path (see \autoref{sec:SI:Dingle}) \cite{Kliemt2016}. The $4f$ electrons of the $^{13+}$Yb atom are subject to crystal effective field (CEF) splitting in which the ground state is predominantly of $|j=7/2, j_z=\pm5/2\rangle$ character \cite{Huesges2018}. The CEF determines the anisotropy of the $4f$ moments favouring an easy magnetic axis along [001] whilst the ferromagnetic phase has moments ordered in the basal plane \cite{Steppke2013}. The currently best-known crystal electric field (CEF) level scheme in the literature places the excited states at \SI{8.5}{\milli\electronvolt}, \SI{12.5}{\milli\electronvolt}, and \SI{30}{\milli\electronvolt} above the ground state \cite{Huesges2018}.

Below the Kondo temperature $\TK\approx\SI{8}{\kelvin}$, strong hybridisation leads to the formation of heavy quasiparticles. Below the Curie temperature $\TC=\SI{170}{\milli\kelvin}$, \YNP\ forms ferromagnetic order \cite{Krellner2011}. In high magnetic fields, a number of Lifshitz transitions (LTs) have been identified from transport and thermodynamic measurements at magnetic fields \Bj\ as listed in \autoref{tab:SI:LT} \cite{Pfau2017}. 
Indeed, heavy-fermion compounds are prone to LTs as the Zeeman energy can quickly become significant when compared to the flat dispersion of quasiparticle bands as observed in CeRu$_2$Si$_2$ \cite{Daou2006} and \YRS\ \cite{Pfau2013}.
In \YNP, the LTs shift to larger fields as the magnetic field is rotated away from the magnetically easy direction [001] thus following the magnetic anisotropy of the material \cite{Karbassi2018}.

We use quantum oscillation studies to understand the evolution of the Fermi surface and quasiparticle properties in high magnetic fields. 
The frequency of the oscillations is proportional to the cross-sectional area of extremal orbits perpendicular to the magnetic field. Through rotation studies, the topology of the Fermi surface can be mapped \cite{Friedemann2013b}. From the field-dependence, the mean free path and potential field-induced changes to the Fermi surface can be detected \cite{Rourke2009}. The temperature dependence of the amplitude allows to extract the quasiparticle mass \cite{Semeniuk2023a}.

We detect a large number of quantum oscillation frequencies in the magnetoresistance $R(B)$ between \SI{6}{\tesla} and \SI{35}{\tesla}. Oscillations are visible both in $R(B)$ and the derivative $\dd R/\dd B$ where the latter is often better suited to subtract the underlying background before using a Fourier transform to identify quantum oscillation frequencies as presented in \autoref{fig:QO}. Prominent frequencies include the highest fundamental frequencies $\Fap=\SI{5.5}{\kT}$ and $\Fap=\SI{4.1}{\kT}$ at intermediate polar angles $\SI{20}{\degree}\leq\theta\leq\SI{60}{\degree}$ (blue traces in \autoref{fig:QO}(a) and (b)), $\Fb=\SI{1.6}{\kT}$ for fields along [100] extending to polar angles $\theta\geq\SI{50}{\degree}$ (orange traces in \autoref{fig:QO}(a) and (b)) and  $\Fc=\SI{0.5}{\kT}$ and $\Fd=\SI{0.4}{\kT}$ for fields along [001] extending to polar angles $\theta\leq\SI{65}{\degree}$ (see also \autoref{fig:SI:Fd_Mass}).
We first focus on the de-renormalisation of quasiparticles at intermediate fields in \autoref{sec:DeRen} before presenting the Fermi-surface reconstruction detected at high fields in \autoref{sec:FSrecon} and discussing scenarios for the Fermi-surface reconstruction in \autoref{sec:FSrecon}. Based on the understanding of the Fermi-surface reconstruction, we are able to compare the angular dependency of quantum oscillations with density-functional-theory (DFT) calculations and draw conclusions on the topology of the Fermi surface in \autoref{sec:FermiSurfaceTopology}.

\section{De-renormalisation of quasiparticles}
\label{sec:DeRen}
\begin{figure}%
\includegraphics[width=\figurewidth]{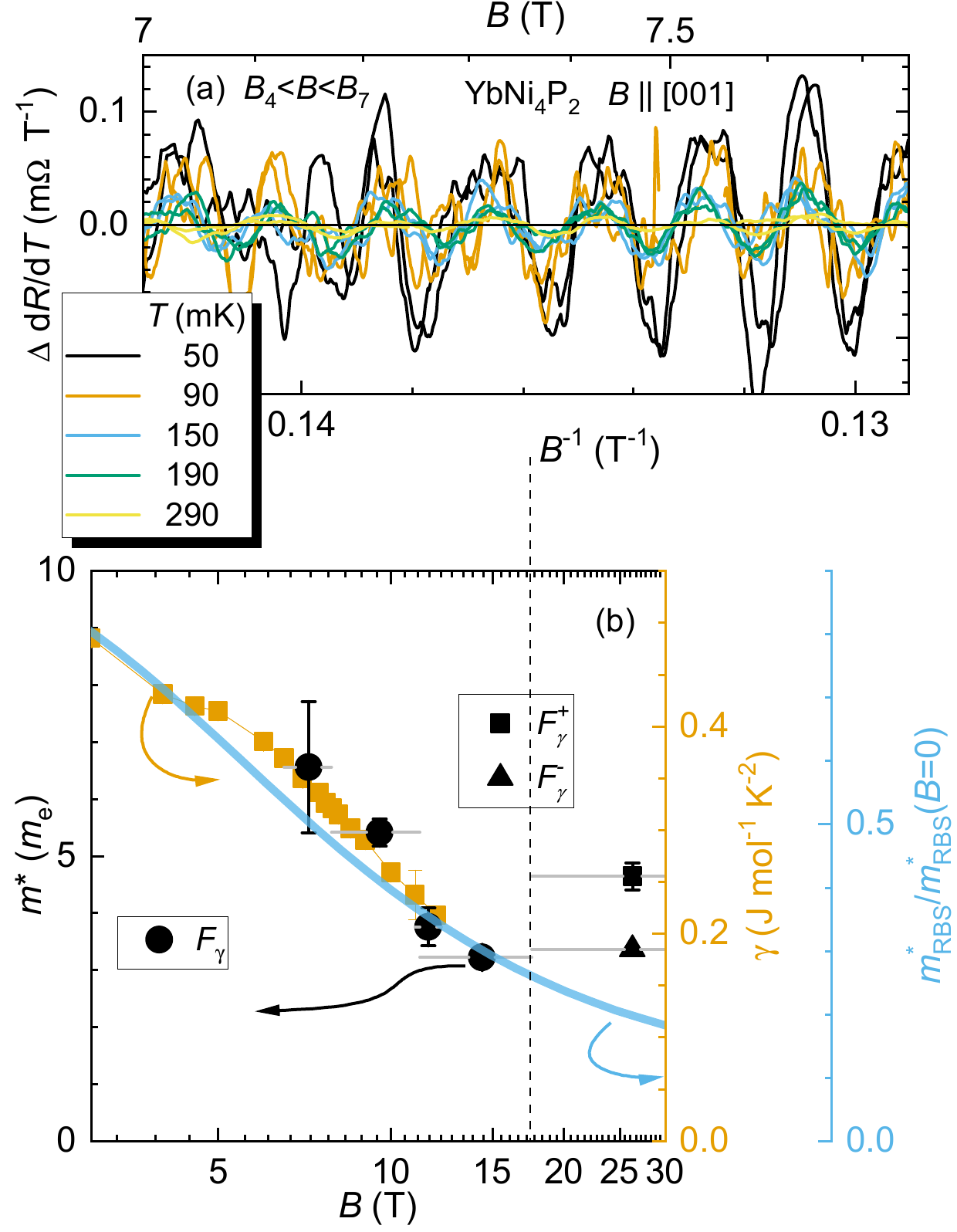}
\caption{\textbf{Field-dependence of quasiparticle mass.} (a) Background-subtracted derivative of the resistance in the field range $\Biv<B<\Bvii$ (cf. \autoref{tab:SI:LT}). 
(b)  Effective masses extracted from Lifshitz-Kosevich (LK) fits (\autoref{fig:SI:QO_DL}) as a function of magnetic field for \Fc\ below \Bix\ and for \Fcpm\ above \Bix. Horizontal (grey) bars denote the field range over which the mass was extracted. Vertical bars denote standard errors from LK fits. Yellow squares show the field dependence of the Sommerfeld coefficient \cite{Pfau2017} on separate axis (right). Solid blue line shows the de-renormalisation of the quasiparticle mass for a single-ion Kondo model with the parameters relevant for \YNP\ (see text for details).}%
\label{fig:MvsH}%
\end{figure}

With increasing magnetic field, we observe a continuous decrease of the quasiparticle effective mass, \mstar\ in \YNP. Effective masses are determined for each individual orbit from the temperature dependence of the oscillation amplitude as shown in \autoref{fig:QO}(c) and \autoref{fig:SI:Fd_Mass}. The decrease of \mstar\ is best seen for the orbit associated with \Fc: the effective mass \mc\ decreases by a factor $\approx2$ between \SIrange{6}{17}{\tesla} as shown in \autoref{fig:MvsH}(b). For the study of the field-dependent effective mass, we focus on \Fc\ because this frequency can be traced over a wide field range from \SIrange{6}{35}{\tesla} owing to the large amplitude (probably as a result of favourable Fermi surface geometry and low Dingle damping (cf. \autoref{sec:SI:Dingle}). We analyse the mass for discrete field ranges separated by the LTs (cf. \autoref{tab:SI:LT} and \cite{Pfau2017,Karbassi2018}). Below the LT at $\Bix=\SI{17.5}{\tesla}$, we observe a single frequency \Fc\ - the splitting of this frequency above \Bix\ will be discussed in \autoref{sec:FSrecon}. The effective mass decreases from $\mc=\SI{7(1)}{\me}$ at \SI{7}{\tesla} to $\mc=\SI{3.2(1)}{\me}$ at \SI{15}{\tesla}. Beyond \SI{17}{\tesla}, i.e. beyond the LT at \Bix, we find different masses for the spin-split branches of \Fcpm\ at $\mstar_{\gamma^{+}}=\SI{4.9(5)}{\me}$ and $\mstar_{\gamma^{-}}=\SI{3.5(6)}{\me}$ (see below for discussion of spin-splitting).

The decrease of \mc\ below \Bix\ reflects the bulk behaviour of all electronic states as it matches the evolution of the Sommerfeld coefficient, \gammaS. The qualitative agreement of \mc\ and \gammaS\ can be seen from \autoref{fig:MvsH}(b) with the same relative decrease in both quantities. Approximate quantitative agreement is apparent from a similar enhancement of both quantities over the non-correlated band-structure calculated with DFT (see \autoref{sec:SI:DFT}): \gammaS\ is enhanced 40 fold over the band mass at \SI{10}{\tesla} while the effective mass of the $\gamma$ orbit is enhanced approximately tenfold, $\mc/\mDFTc\approx10$ over the bare band mass \mDFT\ calculated with DFT. This suggests other bands might be even more strongly renormalised consistent with mass enhancements between 5 and 10 present at much larger fields for \Fa\ and \Fb\ expected to result in masses $\mab/\mDFTab\gtrsim20$ at \SI{10}{\tesla}. A similar decrease in \mstar\ in high magnetic fields has been observed in the Sommerfeld coefficient \cite{Gegenwart2006}, magnetisation \cite{Pfau2017}, and $1/T_1$ NMR relaxation \cite{Bruening2010} in other heavy-fermion compounds but not previously in the quantum-oscillation effective mass. The large field range of the quantum oscillations in \YNP\ affords a direct comparison with theory. 

The decrease of \mc\ and \gammaS\ is well described by the single-ion Kondo impurity model as can be seen from the good qualitative agreement between our experimental data and model data (blue line in \autoref{fig:MvsH}(b)). We use the expectation value of the Zeeman splitting in the CEF ground state and Kondo temperature of \YNP\ to model the evolution of the renormalised effective mass. For a single Kondo ion with valence close to unity, the field-dependent de-renormalisation is conveniently calculated by means of Renormalised Perturbation Theory (RPT) \cite{Edwards2013}. In \autoref{fig:MvsH}, we show that RPT captures well the experimental behaviour for fields above $\approx\SI{2}{\tesla}$. For lower fields, the heat capacity shows a stronger increase and we attribute this to collective effects of the coherent Kondo lattice.


The continuous decrease of \mstar\ shows directly that the quasiparticles become continuously de-renormalised as the Kondo singlet formation is suppressed in high magnetic fields up to \SI{17.5}{\tesla}, i.e. up to the LT at \Bix. At \Bix, however, we observe a drastic change of the Fermi surface topology through a non-linear spin-splitting (see below) which disrupts the continuous de-renormalisation and also leads to an abrupt change of the quasiparticle mass. This behaviour is not captured by the single-ion Kondo model and will require more sophisticated theory in future work. Here, we identify some of the characteristics of the transition at \Bix.

\section{Frequency splitting at \Bix}
\label{sec:FSrecon}
\begin{figure}
    \centering
    \includegraphics[width=\figurewidth]{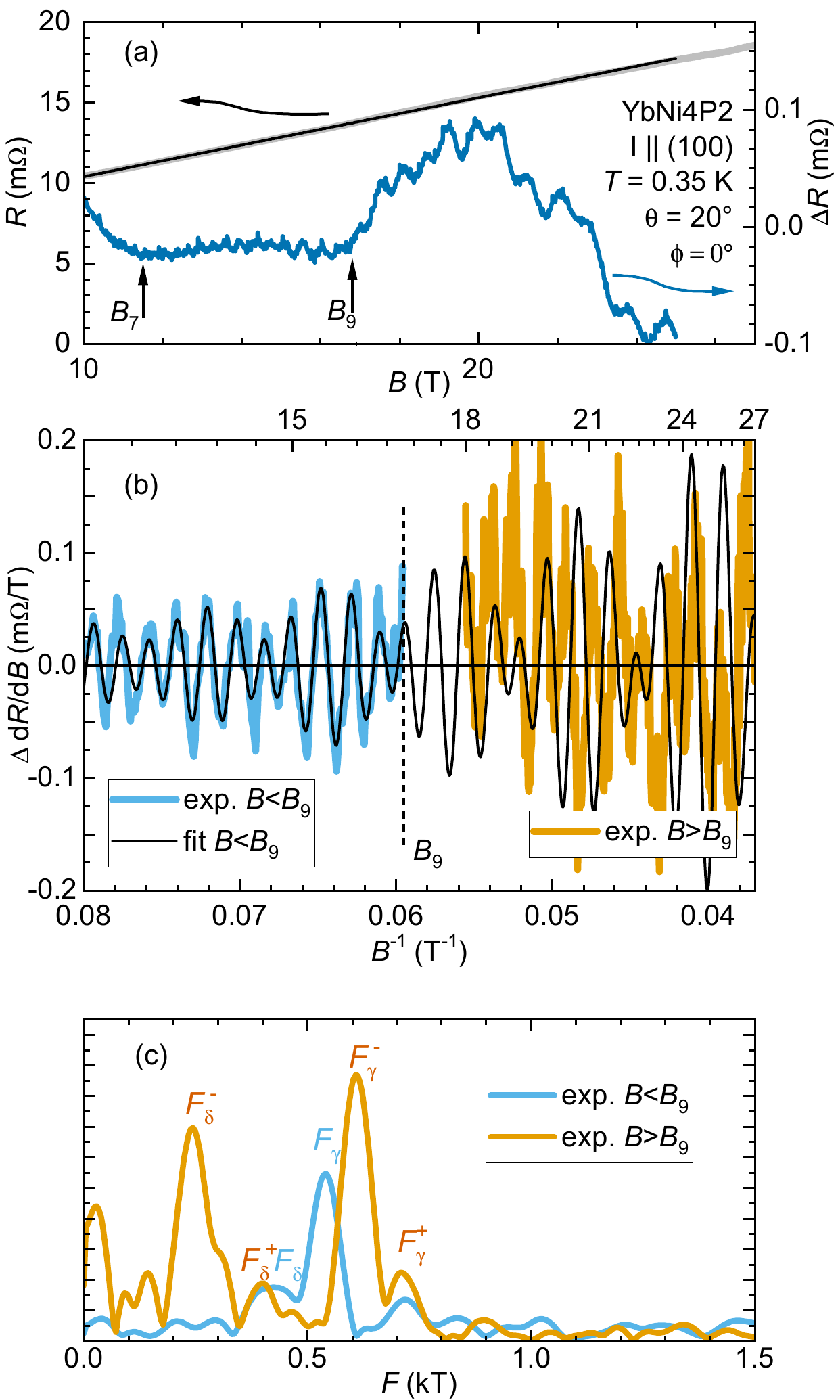}
    \caption{
    \textbf{Change of quantum-oscillation frequencies across \Bix.}
    (a)  Resistance of sample 1 (grey) and linear fit to data (solid black line), left scale. Residual (blue) from linear fit on right scale. The LTs \Bvii\ and \Bix\ (arrows, cf. \autoref{tab:SI:LT}) are visible as kinks in the residual \cite{Karbassi2018,Pfau2017}. 
    (b) Derivative of resistance after subtraction of linear background separately for field range below (blue) and above (yellow) \Bix. Solid black line marks a fit of two quantum oscillations frequencies to the experimental data below \Bix. The fit is extrapolated to above \Bix.
    (c) FFT for derivative below (blue) and above (yellow) \Bix. Labels indicate assigned frequencies.
    }
    \label{fig:QO_LTb}
\end{figure}

We observe drastic changes of quantum oscillation frequencies at \Bix. The changes of quantum oscillation frequencies can be seen in \autoref{fig:QO_LTb} where quantum oscillations are fitted well by two frequencies $\Fd\approx\SI{0.4}{\kT}$ and $\Fc\approx\SI{0.53}{\kT}$ for fields below \Bix\ while this fit fails to describe the data above \Bix. Likewise, the FFT exhibits the two frequencies \Fc\ and \Fd\ for $B<\Bix$ but these two are absent for $B>\Bix$. Instead, new frequencies $\Fcp=\SI{0.7}{\kT}$, $\Fcm=\SI{0.6}{\kT}$, $\Fdp=\SI{0.34}{\kT}$, and $\Fdm=\SI{0.23}{\kT}$ are detected in the FFT for fields $B>\Bix$. 
The fact that the original frequencies \Fc\ and \Fd\ disappear demonstrates that the Fermi surface and/or the quasiparticle properties undergo drastic changes at \Bix. 

\begin{figure}%
	\includegraphics[width=\figurewidth]{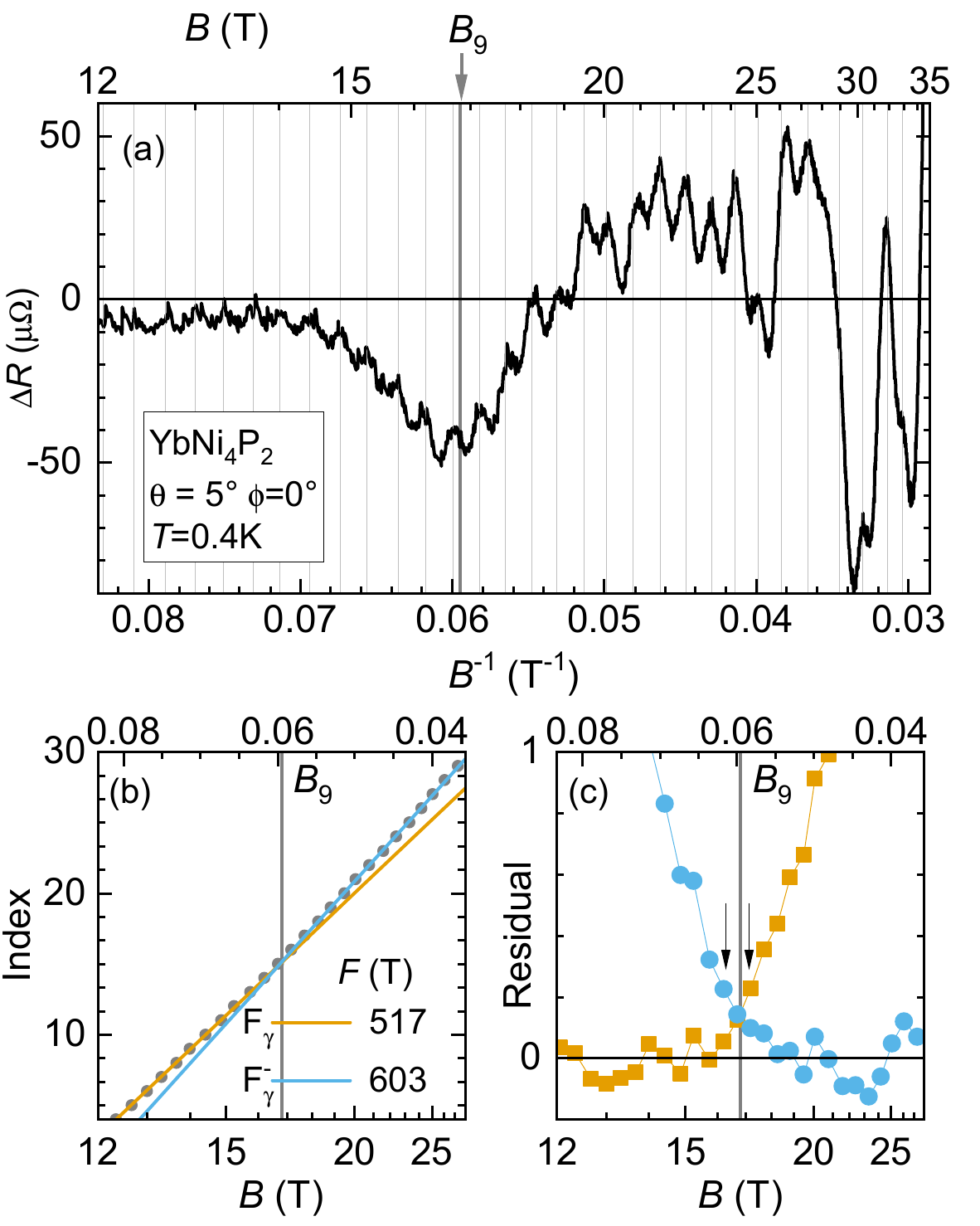}%
\caption{\textbf{Abrupt Change of Quantum Oscillation Frequencies at \Bix.} (a) Resistance after subtraction of 3rd order polynomial. Local maxima in $R(B)$ are indicated by thin grey vertical lines and have been determined from the second derivative. \Bix\ is  indicated as thick grey vertical line. (b) Indices of maxima from (a) vs inverse field together with linear fits for $B<\Bix$ (yellow line) and $B>\Bix$ (blue line). The slope of the lines in (b) corresponds to the quantum oscillation frequencies listed in the figure legend. (c) Residual of linear fits from (b). Arrows in (c) indicate the fields where the residuals deviate from zero. }%
\label{fig:QO_LTa}%
\end{figure}

The changes of quantum oscillation frequencies are sharp at \Bix. This is best seen from the direct indexing of quantum oscillations in \autoref{fig:QO_LTa}. We index maxima (grey vertical lines) from quantum oscillations of the dominant frequencies in $\rho(B)$. The maxima are equally spaced in $R(B^{-1})$ as can be seen from the linear behaviour of maxima positions vs $B^{-1}$ below \Bix\ (yellow line in \autoref{fig:QO_LTa}(b)). The slope in \autoref{fig:QO_LTa}(b) corresponds to $\Fc=\SI{0.52}{\kT}$ for $B<\Bix$. At \Bix, however, the spacing in (a) and thus the  slope in (b) change abruptly to the new frequency $\Fcm=\SI{0.60}{\kT}$ (blue line in \autoref{fig:QO_LTa}(b)). The residuals (\autoref{fig:QO_LTa}(c) highlight that the change from \Fc\ to \Fcm\ happens within a field interval of \SI{0.5}{\tesla} around \Bix\ and within a few oscillations. This abrupt change of quantum oscillation frequencies suggests a sudden change of the quasiparticle properties and/or Fermi surface as discussed below.

\begin{figure}%
	\includegraphics[width=.7\columnwidth]{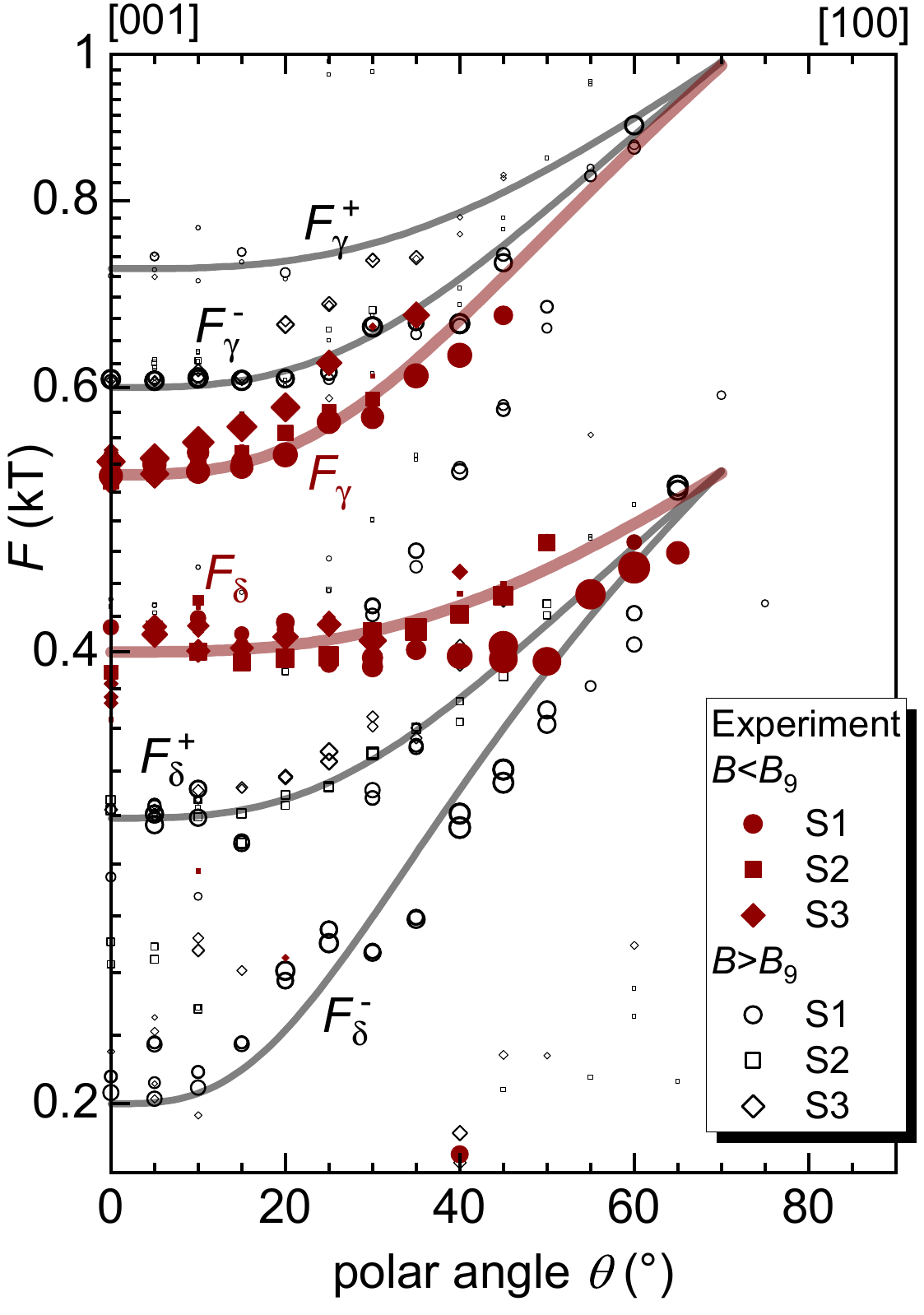}
	\caption{\textbf{Anisotropy of frequency splitting}. (a) Angular dependence of frequencies \Fc\ and \Fd\ detected below \Bix\ but absent above \Bix\ (filled red symbols) together with new frequencies detected above \Bix\ only (open black symbols). Solid lines are guides to the eye for frequencies below (red) and above (grey) \Bix. }
	\label{fig:SpinSplit}%
\end{figure}

The number of detected quantum oscillation frequencies increases significantly above \Bix. While quantum oscillations are always more pronounced at high fields resulting in the detection of additional frequencies, the disappearance of \Fc\ and \Fd\ above \Bix\ combined with the detection of at least double the number of frequencies (\Fcpm, \Fdpm) is not consistent with the standard behaviour of quantum oscillations. Rather, it suggests pair-wise splitting where each low-field frequency (red in \autoref{fig:SpinSplit}) is replaced by two frequencies (black in \autoref{fig:SpinSplit}) above \Bix. This sudden doubling of detected frequencies suggests that a degeneracy (e.g. spin, orbit, or CEF degeneracy) has been lifted above \Bix\ resulting in orbits with different  frequencies. 

A direct observation of frequency-splitting in quantum oscillations is rare. In weakly correlated paramagnets like Pt \cite{Ketterson1970}, MgB$_2$ \cite{Carrington2003}, and ZrTe$_5$ \cite{Wang2018}, a linear spin splitting leads to spin-zeros in the angular dependence of the fundamental quantum oscillation frequency. 
Only in a few materials non-linear spin splitting leads to direct observation of  split frequencies: A splitting of less than \SI{1}{\percent} was observed in CePd$_2$Si$_2$ \cite{Sheikin2004} and in PrBb$_3$ \cite{Endo2002} whilst splitting due to spin-3/2 quasiparticles is suggested for YPtBi \cite{Kim2026a}.

In \YNP, the splitting of the quantum-oscillation frequencies is large compared to other systems displaying spin-split Fermi surface sheets. 
The frequency splitting amounts to $(\Fcp-\Fcm)/\Fc = \SI{20}{\percent}$ and $(\Fdp-\Fdm)/\Fd = \SI{30}{\percent}$ for magnetic field along [001].
A similarly large splitting of frequencies is only reported in \URS, where the $\beta$ frequency is split by $\approx\SI{10}{\percent}$ for fields along the easy axis with the split absent as the field is rotated towards the magnetically hard plane where spin zeros are observed instead \cite{Bastien2019,Aoki2012}. This behaviour is similar to \YNP, where we find that the frequency splitting is also diminished as the field is rotated away from the magnetically easy axis (see below). Large spin splitting is thus linked to the strongly correlated state of \YNP\ and \URS. 
Yet, pronounced differences exist and highlight the complex behaviour of \YNP\ in high magnetic fields: In the case of \URS, the splitting is consistent with a symmetric splitting around the low-field frequency in contrast to the behaviour observed in \YNP\ (see below).

The splitting of the quantum oscillation frequencies is linked to the magnetic anisotropy of \YNP. The change in quantum oscillation frequencies is tied to the transition at \Bix\ and follows the anisotropy of \Bix. This can be seen from the fact that the change of quantum oscillation frequencies occurs at $\Bix(\theta=\SI{5}{\degree})=\SI{16.8}{\tesla}$ for fields close to [001] (see \autoref{fig:QO_LTa}) but shifts to higher fields, e.g. $\Bix(\theta=\SI{45}{\degree})=\SI{19.3}{\tesla}$ (see \autoref{fig:SI:QO_LTc}) in agreement with the shift of the LT as the field is rotated away from [001] (cf. \autoref{sec:SI:QO_LT}). 

The magnitude of the splitting is also tied to the magnetic anisotropy. Largest splitting of quantum oscillation frequencies are detected for fields along [001] as can be seen from \autoref{fig:SpinSplit}. As the field is rotated away from [001], the splitting of the frequencies gradually reduces and no splitting is observed for angles $\theta>\SI{55}{\degree}$. In fact, for fields close to the $ab$ plane, we observe identical frequencies below and above \Bix\ as can be seen for instance for \Fb\ in \autoref{fig:rot_FS}.

In \YNP, the split frequencies at high fields are not symmetric around the non-split frequencies at low fields. In the most simple pictures, split frequencies are associated with separate frequencies for minority and majority spin branches with equal and opposite shift. This is indeed the picture for the split frequencies in CePd$_2$Si$_2$ \cite{Sheikin2004}, PrBb$_3$ \cite{Endo2002}, and \URS\ \cite{Bastien2019,Aoki2012}. In \YNP, we find roughly double the number of frequencies above \Bix\ compared to below \Bix. However, an assignment in pairs that are symmetric around the low-field frequencies is not possible. Instead, we identify pairs of frequencies \Fcpm\ and \Fdpm\ where both members of the pair are above and below the low-field frequencies \Fc\ and \Fd, respectively (cf. \autoref{fig:SpinSplit}). This asymmetric splitting suggests more profound changes to the Fermi surface topology and/or quasiparticle properties.

\section{Models for frequency splitting}
\begin{figure}%
\includegraphics[width=\figurewidth]{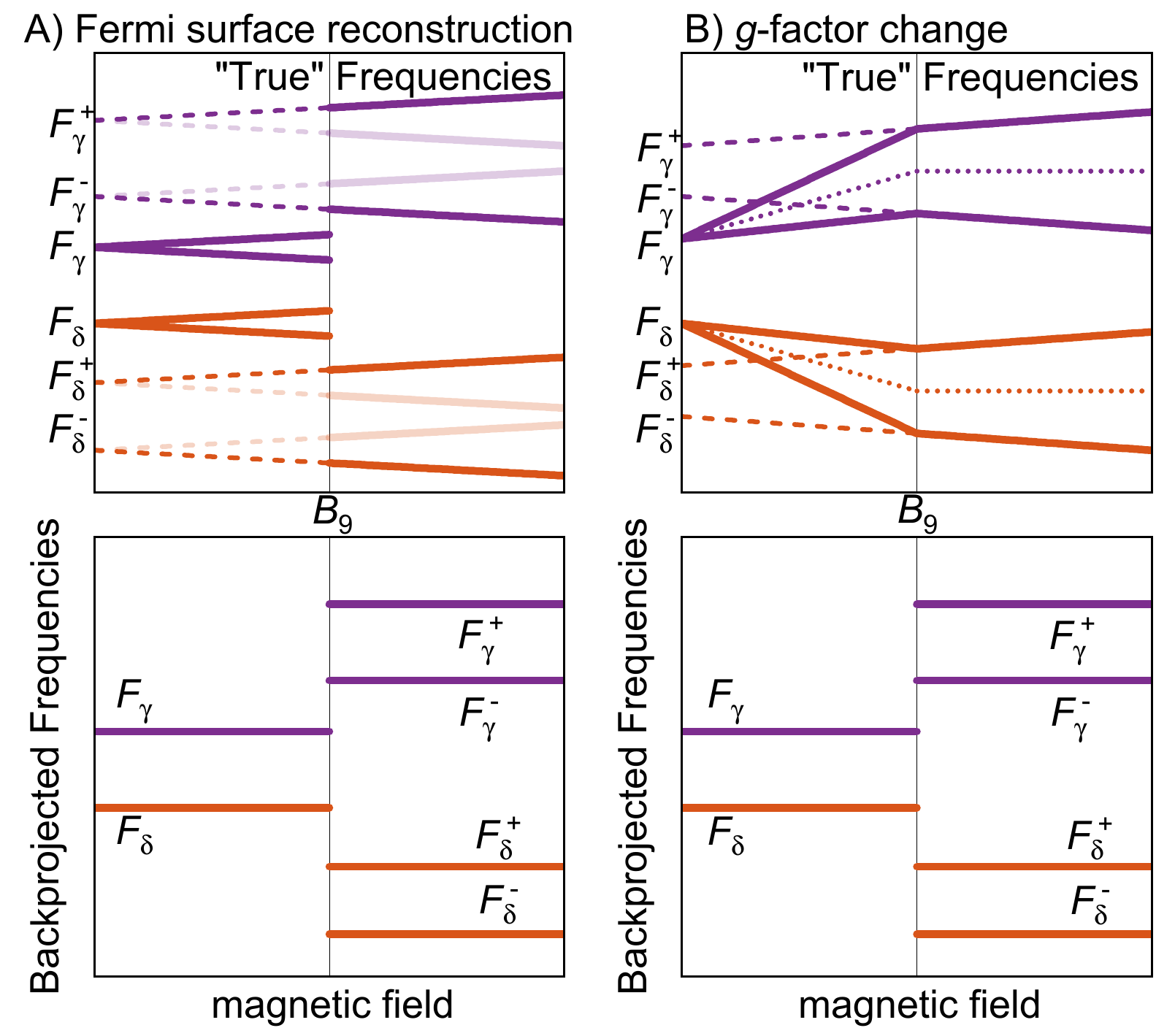}%
\caption{\textbf{Models for the change in quantum oscillation frequencies}. Top panels: The ``true'' frequencies shown as solid lines represent the area of the Fermi-surface cross-section via the Onsager relation. Dashed lines illustrate the back projection. Faint lines in A) highlight the possibility of spin-split frequencies, i.e. only minority and majority spin branches are present (opaque lines). Dotted lines in B) represents the average area between both spin branches including a change associated with the de-renormalisation upon suppression of the Kondo effect in high magnetic fields. Bottom panels summarise the expected frequencies observable in experiment for scenario A) and B).}%
\label{fig:FSscenarios}%
\end{figure}

In order to interpret the observed change of frequencies at \Bix, we need to bear in mind the characteristics of quantum oscillation measurements. Most significantly, quantum oscillations detect the back-projected frequencies 
\begin{equation}
	\Fbp(\Bj) = \FA(\Bj) + B \left(\frac{\dd \FA}{\dd B}\right)_{\Bj}
\label{eq:Fbp}
\end{equation}
which are based on the ``true'' frequency \FA\ proportional to the Fermi-surface cross-section as stated by the Onsager relation and the variation $\dd \FA/\dd B$ of the ``true'' frequency at the field of observation \Bj. Thus, a change in observed quantum oscillations can be caused by changes to one or both terms. However, linear Zeeman splitting into minority and majority branches yields constant frequencies because the field-dependence of the two terms cancel exactly resulting in identical and constant frequencies for both spin branches. This corresponds to the initial behaviour sketched in \autoref{fig:FSscenarios}(A) for fields $B<\Bix$. Thus, only non-linear changes of \FA(\Bj) result in a splitting of observed quantum oscillations.

We sketch two scenarios based on A) sudden jumps of $\FA(B)$ due to a reconstruction of the Fermi surface and B) a sudden change of slope $\dd \FA/\dd B \propto \mstar g$ linked to the product of effective mass and $g$-factor (cf. \autoref{sec:SI:SpinSplit}). We speculate below about the possible driving mechanism for the scenarios.

\subsection{Fermi surface reconstruction}
A reconstruction of the Fermi surface can be used to model the observed changes in scenario A) as sketched in \autoref{fig:FSscenarios}(a). At low fields, the frequencies obey linear spin splitting (constant $g$-factor manifesting as constant opposite slope for spin majority and minority branch) resulting in the observation of a single constant frequency for each orbit \Fc\ and \Fd. The change of observed frequencies is modelled as a Fermi-surface reconstruction at \Bix, i.e. a sudden change to the crossection area of the sheets giving rise to \Fc\ and \Fd. The doubling of the number of frequencies is modelled as a doubling of Fermi surface sheets as a result of lifting a degeneracy of either spin or other degree of freedom or change in symmetry. The back-projected frequencies (intersect of dashed lines) show a sudden jump and double the number of observed frequencies. In experiment, we observe that the frequencies above \Bix\ are not split symmetrically around the low-field frequencies. In scenario A), this is modelled by an overall increase of frequencies \Fcpm\ and reduction of \Fdpm\ and may be a result of opposite band character for the two orbits, e.g. electron- versus hole-like (consistent with the assignment of frequencies to orbits in \autoref{sec:FermiSurfaceTopology}).

A wide range of phenomena can cause non-linear field dependency of the extremal Fermi surface area. This includes spin-split Fermi surfaces in ferromagnets \cite{Lonzarich1974,Hayden1986,Ruitenbeek1982}, altermagnets with spin-orbit coupling \cite{Li2025d}, other phase transitions like antiferromagnetism \cite{Mercure2009}, and metamagnetic transitions as observed in CePd$_2$Si$_2$ \cite{Sheikin2003}, CeRu$_2$Si$_2$ \cite{Meulen1991} and UPt$_3$ \cite{McCollam2020}. In \YNP, we detect the split quantum oscillation frequencies at temperatures above the ferromagnetic transition and in fields perpendicular to the ordered moment ruling out a simple spontaneous ferromagnetic state. The symmetry group of \YNP\ does not allow for altermagnetism. We can also rule out a Pomeranchuk transition as detailed in \autoref{sec:SI:alternatives}. 

We can rule out a simple metamagnetic transition as observed for instance in CeRu$_2$Si$_2$ \cite{Aoki2001a}. For a metamagnetic transition, the observed change in frequency is proportional to the change in area due to a spin polarisation (assuming constant $g$). Given the effective masses and frequencies observed for $\gamma$ and $\delta$ orbits, we can estimate the change in magnetic moment and magnetisation that would correspond to the spin-splitting of the \Fc\ and \Fd\ frequencies  across \Bix. The frequency changes of $\approx\SIrange{20}{30}{\percent}$ for \Fc\ and \Fd\ would correspond to a volume change of \SIrange{30}{50}{\percent}. As we have evidence that all Fermi surfaces are spin-split (see below), we would expect this to yield a change of more than \SI{50}{\percent} of the polarised moment. By contrast, the magnetisation is already within \SI{10}{\percent} of full polarisation at \Bix\ \cite{Kliemt2023}. Indeed, the magnetisation data rule out a metamagnetic transition of more than \SI{5}{\percent} of the full moment of the Yb ground state. This is insufficient to explain the large frequency splitting.

Excluding these mechanisms, highlights that further work is required to understand the high-field electronic structure of \YNP\ and related heavy fermions. Considering excited states of the CEF scheme in high fields is a promising route as indicated by the correlation of the magnitude of the frequency splitting with the magnetic anisotropy.

\subsection{\texorpdfstring{Abrupt change of $g$-factor}{Abrupt change of g-factor}}
As an alternative interpretation of the data, we consider model B) based on a change of the second term in \autoref{eq:Fbp}. Here, we model the jump of the quantum oscillations with a combination of linear Zeeman splitting, an abrupt change of the $g$ factor, and a linear suppression of the Kondo effect as illustrated in \autoref{fig:FSscenarios}(b). Below \Bix, a linear Zeeman effect (constant low-field \gl) gives rise to linear growing (shrinking) Fermi surface area for the majority (minority) branch (splitting of solid lines in \autoref{fig:FSscenarios}(b). 
For fields $B\leq\Bix$, these linear effects result in a single back-projected frequency for each \Fc\ and \Fd\ consistent with our in quantum oscillation measurements (\autoref{fig:QO_LTb} and \autoref{fig:SpinSplit}).
The central assumption of scenario B) is that the Zeeman shift abruptly changes slope at \Bix\ which results in an abrupt splitting of back-projected quantum oscillation frequencies while the Zeeman shift is linear below and above \Bix.

In model B), The form of frequency splitting observed in experiment reveals insight about the de-renormalisation of the Fermi surface from the suppression of the Kondo effect because the back-projections provide a snap-shot of the field evolution up to \Bix. The fact that no field-dependence is observed for \Fc\ and \Fd\ below \Bix\ and no field-dependence is observed for \Fcpm\ and \Fdpm\ above \Bix\ means changes to the true frequencies are linear in both regimes  (dotted line in \autoref{fig:FSscenarios}(b)) and only becomes non-linear at the transition \Bix. The fact that the frequencies observed above \Bix\ are split asymmetrically around the low-field frequencies suggests that the mean Fermi surface has changed up to \Bix. We incorporate a linear reduction of the ``true'' frequencies into our model (dotted line in \autoref{fig:FSscenarios}(b)). We relate this to the reduction of the Fermi-surface volume as the Kondo effect is suppressed in high fields consistent with the reduction of the quasiparticle mass (\autoref{sec:DeRen}).
On top of this, the spin-split branches (solid lines in \autoref{fig:FSscenarios}(b)) of each Fermi surface are modelled with linear Zeeman-splitting (see \autoref{sec:SI:SpinSplit} for details). 

Within scenario B), we deduce an abrupt change in the electronic $g$ factor from the magnitude of the frequency splitting. The change in slope at \Bix\ is set by the change of the product of effective mass and $g$-factor, i.e. $\Delta F= \Fcp-\Fcm \propto \Delta (\mstar g)$ (see \autoref{sec:SI:SpinSplit}). Given that we measure a small difference in effective mass, we estimate the change in $g$-factor to be $\Delta g \approx 0.5(2)$. 
Model calculations predict that the $g$ factor can be enhanced in Kondo systems and may decrease discontinuously. Indeed, an abrupt change in the $g$-factor in magnetic field has been predicted in quantum dots from the interplay of Kondo screening and quantum Hall edge states where the field anisotropy plays an important role \cite{Heine2016}. In \YNP, the evolution of the Fermi surface might contribute further to the discontinuity of $g$. 

From the direction of the asymmetric splitting, we identify electron- and hole-like pockets for the frequencies \Fc\ and \Fd. The asymmetry is different for \Fc\ and \Fd. The split frequencies \Fcpm\ are above \Fc\ whilst \Fdpm\ are below \Fd. This means that \Fc\ is growing whilst \Fd\ is shrinking with the de-renormalisation in fields up to \Bix\ (positive/negative slope on dotted line in \autoref{fig:FSscenarios}(b)). Shrinking (growing) Fermi surface volume is expected for hole-like (electron-like) bands in Yb-compounds as the Kondo effect is suppressed in magnetic field. This is because, in zero field, the Kondo effect has increased (decreased) the size of the hole-like (electron-like) orbits due to the addition of hole-like states from the Yb moments (cf \autoref{fig:SI:Spaghetti_LNP}). The inferred electron (hole)-like behaviour is consistent with the identification of orbits based on their topology in \autoref{sec:FermiSurfaceTopology}.

\begin{figure*}
    \centering
    \includegraphics[width=.535\textwidth]{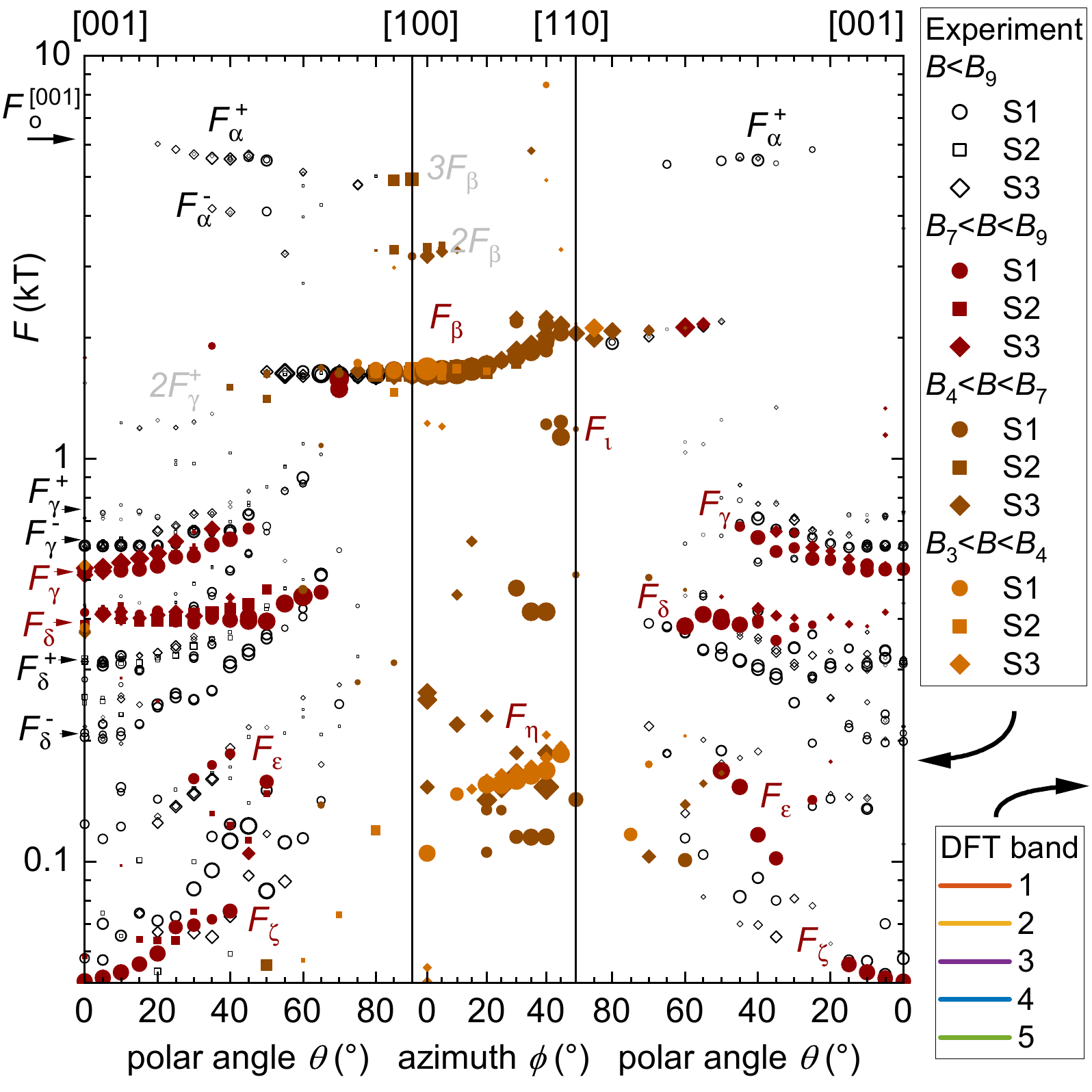}
		\includegraphics[width=.455\textwidth]{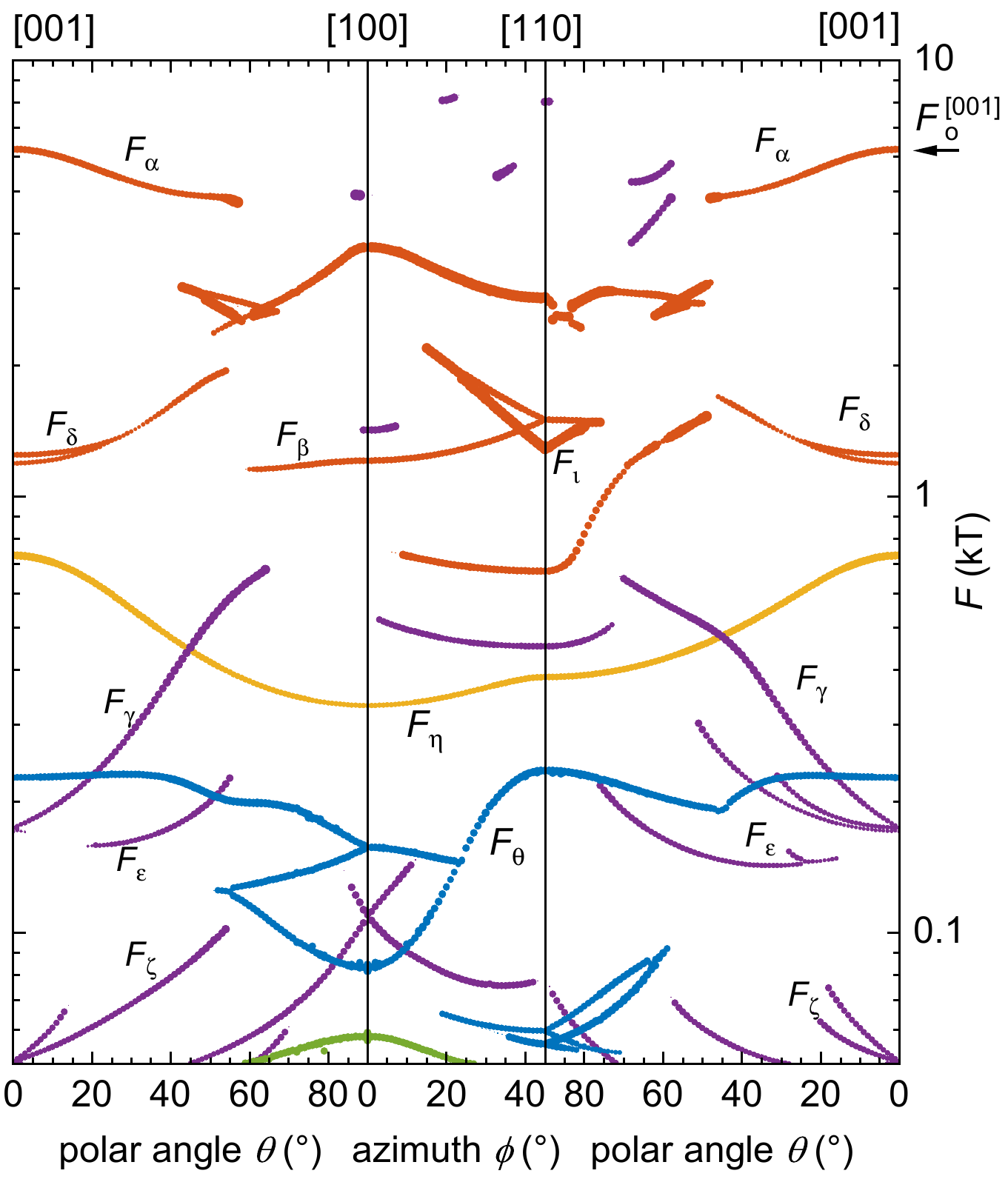}
    \caption{\textbf{Angle dependence of quantum oscillation frequencies.} (a) Experiment below and above \Bix. Symbol size scales with FFT amplitude. (b) DFT $f$-core calculation with band shifts for individual bands corresponding to partial contribution of Yb $f$ states. Symbol size is inversely proportional to curvature factor and thus provides indication of expected signal magnitude. 
		}
    \label{fig:rot_FS}
\end{figure*}

We now turn to a discussion of the anisotropy of the spin splitting.
The frequency splitting is linked with the magnetic anisotropy. Within scenario B), this suggests a reduction of the change $\Delta g$ as the field is rotated away from the [001] direction and $\Delta g = 0$ for $\theta\geq\SI{55}{\degree}$. The link with the magnetic anisotropy suggests that the frequency splitting is caused by changes to the CEF. We speculate that some of the CEF levels that split in magnetic field mix with higher CEF levels and thus induce changes to the ground state and consequently to the electronic structure and quasiparticle properties. More detailed computational work incorporating the correlated state of \YNP\ will be needed to identify how the CEF scheme evolves in field and how it influences the electronic structure.

In order to distinguish between the two models A) and B), other probes will be required.  Electron-spin resonance \cite{Kataev2009} might allow to observe the change of $g$ factor of scenario B) in future work. Further probes sensitive to the Fermi surface and electronic structure are also desirable if they can be implemented at the high fields required to traverse \Bix.
Given the additional work needed to understand the high-field electronic structure, we focus on the low-field frequencies in our comparison with DFT calculations where we do not attempt to model the frequency splitting.

\section{Fermi surface topology}
\label{sec:FermiSurfaceTopology}
\begin{figure}
    \centering
    \includegraphics[width=.6\columnwidth]{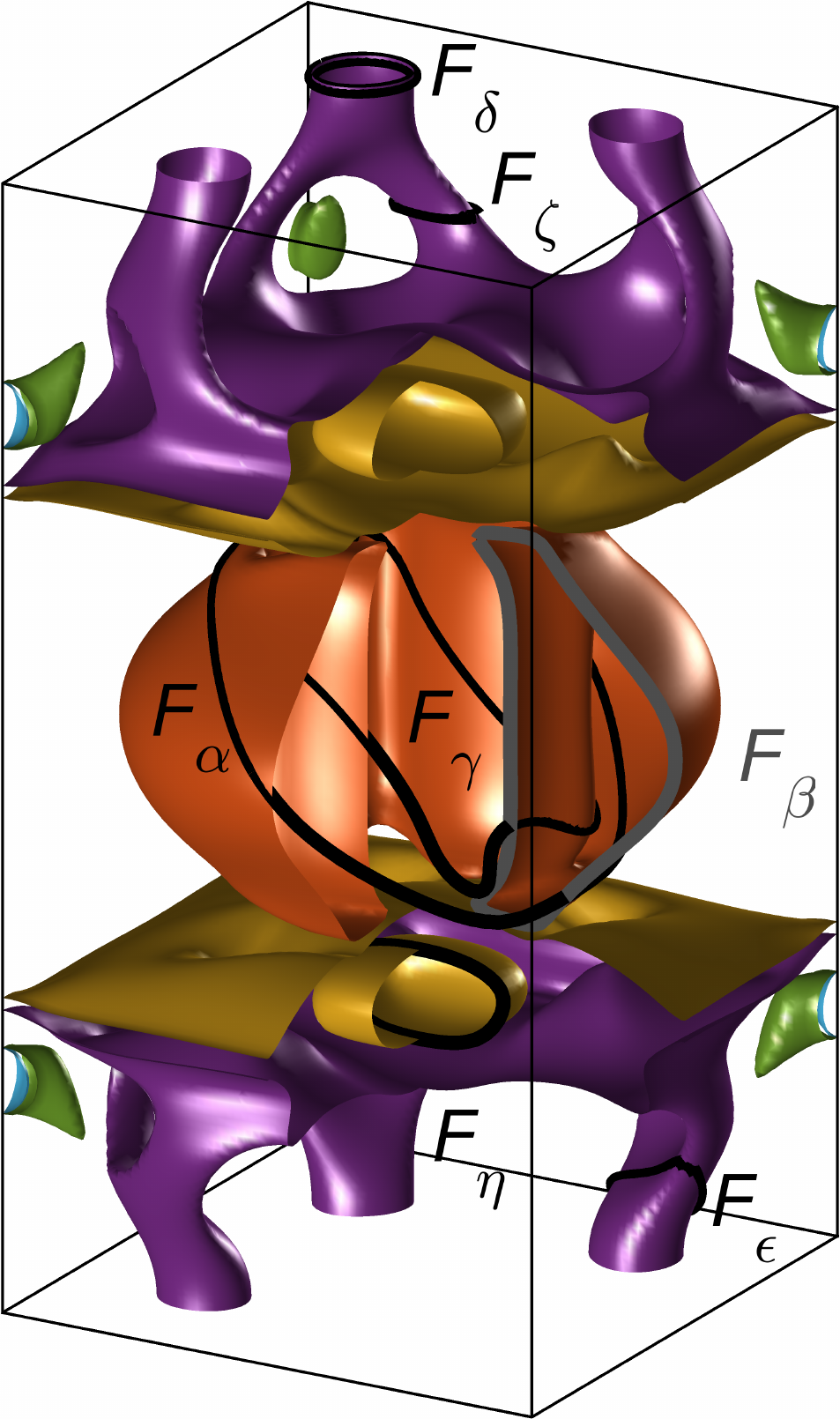}
    \caption{\textbf{$f$-core Fermi surface of \YNP}. DFT calculations with spin-orbit interaction have been used to determine the Fermi surface topology. Small rigid band shifts  have been included to obtain good qualitative agreement with the angular dependence of the quantum experimentally detected oscillation frequencies (cf. \autoref{tab:SI:DFTShifts}). Lines highlight orbits assigned to specific frequency branches.
		}
    \label{fig:FS}
\end{figure}

In order to understand the electronic structure, we model the Fermi surface topology with $f$-core DFT calculations and incorporate the effect of the added $f$ degrees of freedom by shifting the chemical potential (typically a few tens of \unit{\milli\electronvolt}) to obtain agreement with the angular dependence of the detected quantum oscillation frequencies (see \autoref{sec:SI:DFT} for computational details). The resulting Fermi surface (\autoref{fig:FS}) comprises three major sheets: Two flat sheets (yellow) perpendicular to the [001] direction are located half way between $\Gamma$ and $Z$ in agreement with earlier DFT \cite{Krellner2011} and ARPES studies \cite{Dai2025}. A doughnut-shaped Fermi surface (orange) is present around the $\Gamma$ point. A jungle-gym Fermi surface (purple) connects beyond the top/bottom of the Brillouin zone (BZ). Two additional small pockets are located at the edges of the BZ (blue and green). 

We assign Fermi surface sheets to observed quantum oscillations through their angular dependence. 
In our comparison of observed and predicted frequencies (\autoref{fig:rot_FS} (a) and (b)), we focus on the non-spin-split frequencies detected below \Bix\ (red and orange filled symbols in \autoref{fig:rot_FS}(a)) as we do not model the spin-splitting effect in DFT calculations. 
We observe good agreement with predicted frequencies from the doughnut-shaped Fermi surface (band 1, orange), and the jungle-gym (band 3, purple) at the top of the BZ and we find frequencies that match the small spheroids of the predominantly 1D band (band 2, yellow). 

The frequency \Fb\ is best matched to the vertical cross-section of the doughnut Fermi surface (\autoref{fig:FS}). The angular dependence of the observed frequency matches the prediction very well including the kink close to [110].  The fact that \Fb\ is observed over the full rotation between [100] and [110] proves that the doughnut has no necks along the $X$ point. 
The magnitude of the observed frequency is larger than the prediction by DFT. Together with the outer diameter being limited by the absence of necks and size of the Brillouin zone, this suggests that the inner diameter of the doughnut is smaller than predicted by DFT.

The frequency \Fc\ is associated with the inner cylinder of the doughnut although the smaller frequency in experiment indicates that DFT overestimates the size of the hole (consistent with the fact that \Fb\ is larger than predicted by DFT). The magnitude of the observed frequency \Fc\ is in good agreement with ARPES measurements which showed a small hole in the doughnut in zero magnetic field for energies $E=\EF-\SI{80}{\milli\electronvolt}$.

The frequencies \Fd, \Fe, and \Ff\ are assigned to the cylindrical connections in the jungle-gym Fermi surface. The fact that \Fd\ is larger than predicted by DFT suggests that these cylinders are larger in diameter. The larger diameter also implies that the minor bands 4 and 5 are pushed to higher energy and their Fermi surface either reduced in their size or potentially absent at \EF. 

From the asymmetric spin-split frequencies \Fcpm\ and \Fdpm, we concluded earlier that \Fc\ and \Fd\ would be orbits on electron-like and hole-like Fermi surfaces, respectively. With the mapping of the angular dependence between experiment and theory, this is corroborated. Indeed, \Fdpm\ are now identified as the spin-split orbits of the necks (\Fd) in the jungle-gym. These originate  from a hole-like band (purple band in \autoref{fig:SI:Spaghetti_LNP} between $Z$ and $A$). This orbit is  expected to shrink as the Fermi surface becomes ``smaller'' with less Yb hole states contributing. 
\Fcpm\ are identified with the spin-split orbits of the inner tube (\Fc) in the doughnut. This Fermi surface is electron like and the orbit is expected to increase as the Fermi surface becomes ``smaller'' (cf. orange band in \autoref{fig:SI:Spaghetti_LNP} between $\Gamma$ and $X$).

The largest detected frequency \Fap\ qualitatively matches the predicted angular dependence for the belly orbit of the doughnut-shaped Fermi surface around $\Gamma$ (\autoref{fig:FS}). In addition, we observe a second frequency \Fam\ close by. Given that both frequencies are detected for fields above \Bix, they are likely spin-split frequencies of \Fa\ which itself is not detected below \Bix\ (likely due to the additional damping of high frequencies). This gives rise to the possibility of \Fap\ forming necks even if \Fa\ would not have necks at low fields $B<\Bix$. Indeed, \Fap\ reaches the maximum possible frequency $\Fsc=\SI{6.2}{\kilo\tesla}$ for a circular orbit around the belly for $B\parallel[001]$ before disappearing for $\theta\leq\SI{25}{\degree}$. This is exactly what would be expected for a belly orbit on a Fermi surface with necks. By contrast the smaller frequency \Fam\ corresponds to a substantially smaller Fermi surface consistent with the absence of necks. Thus, even though \Fap\ likely forms necks, the doublet \Fapm\ together with the analysis of \Fb\ above suggests that the low-field Fermi surface has no necks on the doughnut. The formation of necks on the doughnut might be a driving factor for the electronic transition at \Bix\ and the resulting change to the Fermi surface topology and quasiparticle properties.

The frequency $\Fh=\SI{160}{\tesla}$ observed for fields in the basal plane matches the prediction of the ellipsoid on band 2. Whilst the flat sheets of band 2 cannot be detected in quantum oscillations due to the absence of extremal orbits, our observation is consistent with earlier ARPES studies which observed sheets and ellipsoids but could not establish their separation \cite{Dai2025}. Our quantum oscillation measurements indicate that this ellipsoid is separated from the sheets.

\section{Conclusion}
In summary, we detect quantum oscillations from all major Fermi surfaces. Our combined quantum oscillation and DFT study provides evidence that the Fermi surface as shown in \autoref{fig:FS} is indeed realised in \YNP\ in the heavy-fermion state with small modifications including a decreased inner tube inside the doughnut. Thus, our study complements earlier ARPES measurements which directly detected the 1D sheets. Indeed, through the back-projection of quantum oscillations, we detect the ``large'' Fermi surface at fields $B<\Bix$ consistent with ARPES studies. This ``large'' Fermi surface forms the starting point for further work to understand the peculiar high-field properties of \YNP\ including the de-renormalisation of quasiparticle mass and shrinking of the Fermi-surface volume up to \Bix\ and the abrupt change of the electronic structure and/or quasiparticle characteristics inferred from the change in quantum oscillation frequencies.

\begin{acknowledgments}
The authors are grateful for discussions with S. Julian and J. Sourd. This work was partially supported by the EPSRC under grants EP/R011141/1, EP/L025736/1, EP/N026691/1, EP/L015544/1. 
\end{acknowledgments}

\section*{Additional information}

Data are available at the University of Bristol data repository \cite{YNP_Data}.


%

\clearpage
\section*{Supplementary Information}
\clearpage

\renewcommand{\thefigure}{S\arabic{figure}}
\renewcommand{\thetable}{S\arabic{table}}
\renewcommand{\thesection}{S \Roman{section}}
\renewcommand{\theequation}{S\arabic{equation}}
\setcounter{figure}{0}
\setcounter{table}{0}
\setcounter{section}{0}
\setcounter{equation}{0}


\section{Experimental Details}
\label{sec:SI:Exp}
Samples were synthesised using the Czochralski method as described previously \cite{Kliemt2016}. Samples were cut from an oriented single-crystal and optimised for high-sensitivity resistance measurements, typically \SI{1}{\milli\meter} long, with a cross-section $\approx 0.2 \times 0.2 \si{\milli\meter\squared}$. Different samples with current along [100] and [001] where measured and both longitudinal and transverse magnetoresistance showed quantum oscillations. Details for all samples are listed in \autoref{tab:SI:samples}.

\begin{table}%
	\begin{tabular}{ll}
		\hline
		\hline
		sample & $I$ \\
		\hline
		S1 & [100] \\
		S2 & [001] \\
		S3 & [100] \\
		\hline
		\hline
	\end{tabular}
	\caption{Samples used in this study and direction of electrical current.}
	\label{tab:SI:samples}
\end{table}%

The samples used in this study were characterised previously and showed clear signatures in magnetoresistance at several of the LTs as listed in \autoref{tab:SI:LT}. The angular dependence of these LTs $B_i(\theta)$ has been previously mapped \cite{Karbassi2018} and is used here to analyse the quantum oscillations in field ranges demarcated by the visible LTs. We assume isotropic behaviour  in the $ab$ plane, i.e. $B_i(\phi,\theta)=B_i(\theta)$. 

\begin{table}%
	\begin{tabular}{lllll}
	\hline
	\hline
		\Biii & \Biv  & \Bvii & \Bviii & \Bix \\
		\hline
		4.8   & 5.1   & 7.8   & 11    & 16.8 \\
		\hline
		\hline
	\end{tabular}%
	\caption{Lifshitz transitions detected in the samples of \YNP\ used in this study and previous studies on the same samples for fields along [001]. }
	\label{tab:SI:LT}
\end{table}%

Measurements were conducted at the high-field laboratory, HFML, at fields up to \SI{35}{\tesla} in a $^3$He cryostat with a base temperature of $\approx\SI{0.3}{\kelvin}$ and at the University of Bristol in a $^3$He/$^4$He dilution refrigerator in fields up to \SI{12}{\tesla}. Standard 4-probe AC methods were used to record the magnetoresistance. Both the magnetoresistance, $R(B)$ and the derivative $\dd R / \dd B$ were analysed for quantum oscillations to gain sensitivity to low and high frequencies, respectively and to aid background subtraction. Polynomial fits were used to subtract the background where we checked carefully to avoid introducing artefacts in the subsequent Fourier transformation. 

The angular dependence of quantum oscillation frequencies was mapped using a single-axis rotator where samples were aligned such that the magnetic field traced trajectories [001]-[100], [100]-[110], and [110]-[001]. The initial alignment of the samples had an uncertainty $\approx\SI{5}{\degree}$ in all directions. 
The polar angle $\theta$ and the azimuthal angle $\phi$ measure the angle between the magnetic field and the [001] direction and the projected angle in the [100]-[010] away from [100], respectively (see inset in \autoref{fig:QO} (a)).


\section{Dingle Analysis of quantum oscillations}
\label{sec:SI:Dingle}

The mean free path has been extracted for quantum oscillations that are present over a sufficiently large field range and that can be isolated with the help of Fourier filters. Through fits of the Dingle factor as illustrated by the gray line in \autoref{fig:QO}, we determine the mean free path. These range from \SI{50}{\nm} to  \SI{100}{\nm} and are listed in \autoref{tab:SI:QO}.

\begin{table*}%
\setlength{\tabcolsep}{12pt}

\begin{tabular}{lllllll|l}
\hline
\hline
    orbit & $F(\unit{\tesla})$ & $m^{\star} (\unit{\me})$ & $l_0 (\unit{\nano\meter})$ & $\theta (\unit{\degree})$ & $\phi (\unit{\degree})$ & field range & $\mDFT (\unit{\me})$ \\
\hline
    \Fap  & 5510  & 6.1(5) & 90(10) & 49    & 0     & $\Bix<B$ &  \\
    \Fam  & 4140  & 10(1) & 70(20) & 49    & 0     & $\Bix<B$ &  \\
    \Fa   &       &       &       &       &       &       & 1.3 \\
    \Fb   & 1640  & 5.8(4) & 51(5) & 90    & 0     & $\Bvii<B<\Bviii$ & 1 \\
    \Fc   & 530   & 3.2(1) & 130(30) & 0     & -     & $\Bviii<B<\Bix$ & 0.5 \\
    \Fcp  & 720   & 4.9(5) &       & 0     & -     & $\Bix<B$ &  \\
    \Fcm  & 614   & 3.5(6) &       & 0     & -     & $\Bix<B$ &  \\
    \Fd   & 373   & 12.2(3) & 110(30) & 0     & -     & $\Bvii<B<\Bviii$ & 1.3 \\
    \Fdp  & 318   & 2.9(2) &       & 0     & -     & $\Bix<B$ &  \\
    \Fdm  & 210   & 3.5(6) &       & 0     & -     & $\Bix<B$ &  \\
    \Fe   & 120   & 1.7(1) &       & 0     & -     & $\Bix<B$ & 0.6 \\
    \Ff   & 50    & 1.5(2) &       & 0     & 0     & $\Bviii<B<\Bix$ & 0.35 \\
    \Fg   & 160   &       &       & 90    & 35    & $\Biii<B<\Biv$ & 0.35 \\
    \Fh   & 430   & 5.2(6) &       & 90    & 0     & $\Bix<B$ & 1.2 \\
    \Fi   & 1160  & 9(1)  &       & 90    & 45    & $\Bvii<B<\Bviii$ & 1.5 \\
\hline
\hline
\end{tabular}%
	\caption{\textbf{Summary of quantum oscillation frequencies and cyclotron masses.} Effective mass \mstar\ is extracted by fitting the \LK\ temperature dependence to the FFT peak amplitude. Mean free path $l_0$ is extracted by directly fitting damped sinusoids to dominant frequencies. Band masses \mDFT\ are obtained from the $f$-core calculations for the individual orbits with SKEAF \cite{Rourke2012}.
}
	\label{tab:SI:QO}
\end{table*}%

\section{Effective mass analysis}

\begin{figure}%
\includegraphics[width=\figurewidth]{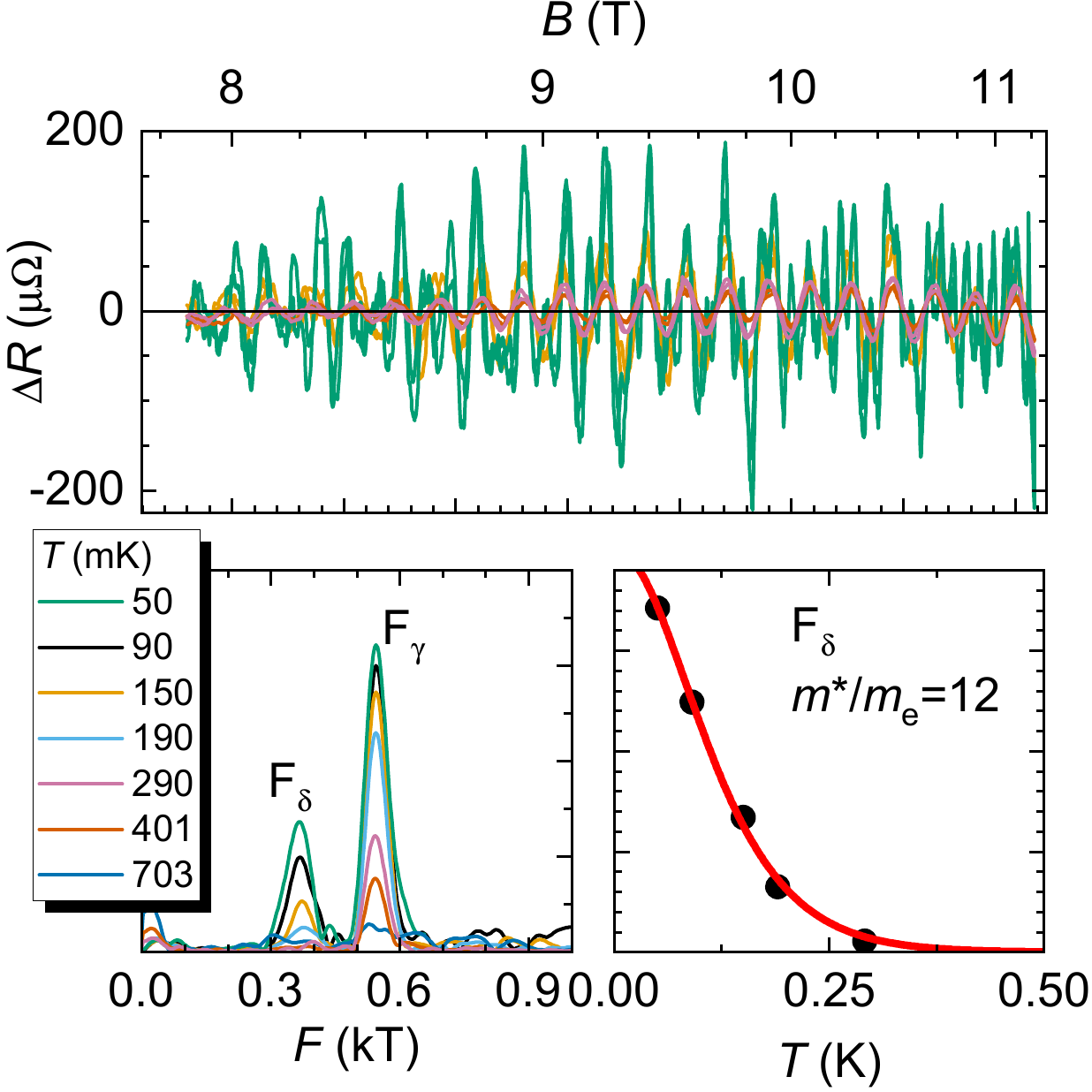}%
\caption{\textbf{Quasiparticle effective mass.} (a) Resistance after subtraction of 1st order polynomial for fields $\Bvii<B<\Bviii$ applied along the [001] direction. (b) FFT of the resistance data in (a). (c) Temperature dependence of the amplitude of the \Fd\ frequency. Line denotes a LK fit.}%
\label{fig:SI:Fd_Mass}%
\end{figure}

\begin{figure}%
	\includegraphics[width=\figurewidth]{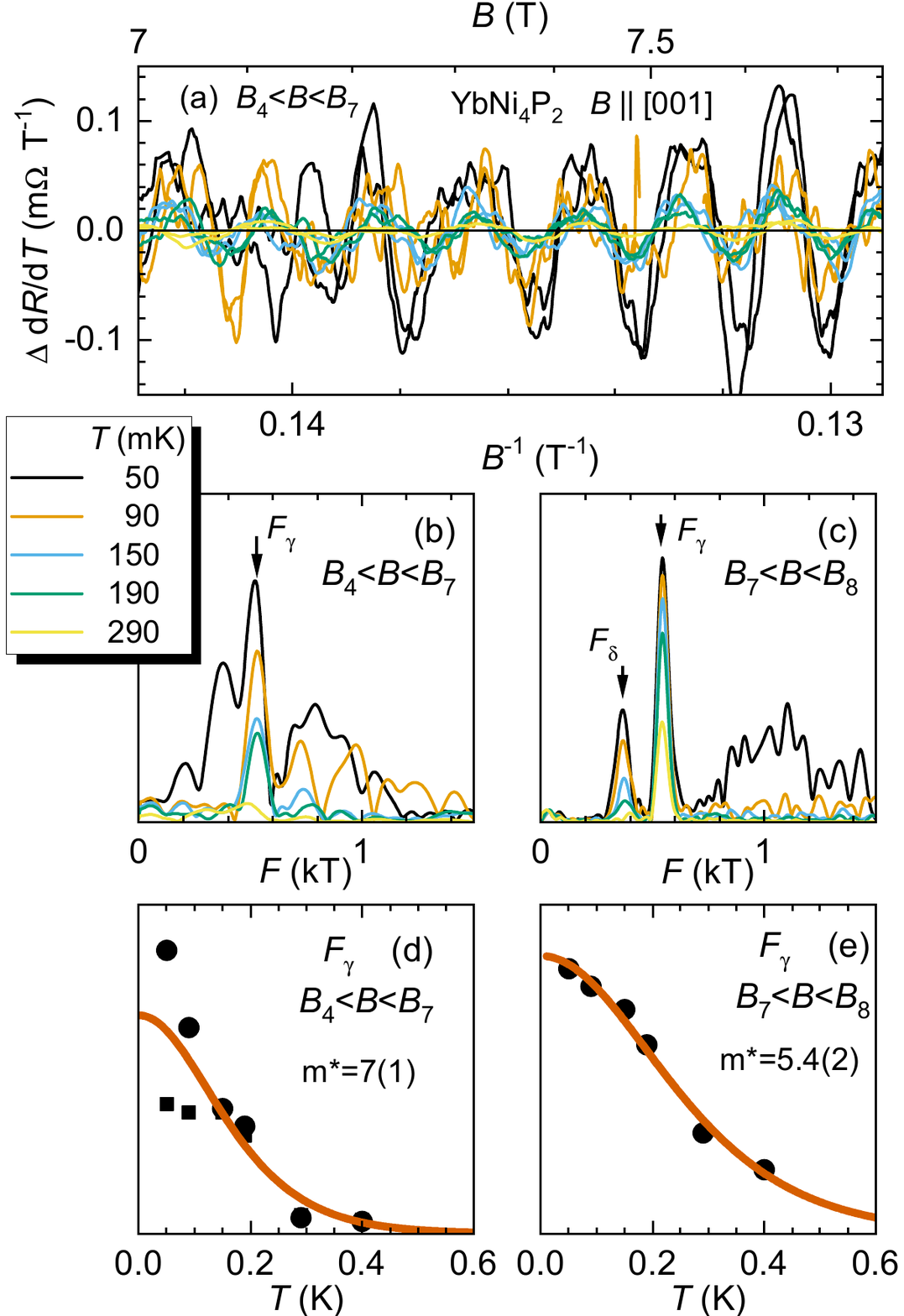}%
	\caption{\textbf{Field-Dependence of quasiparticle mass.} (a) Background-subtracted derivative of the resistance. The data at lowest temperatures have been recorded using a reduced excitation current leading to reduced signal-to-noise ratio. (b) and (c) Fourier transforms of the background subtracted data for different field ranges demarcated by LTs \Biv, \Bvii, and \Bviii. (d) and (e) Temperature dependence of the peak amplitude for the different field ranges shown in (b) and (c). Solid lines in (d) and (e) mark Lifshitz-Kosevich fits to the data. In (d), results for two different field polarities are included (circles and squares).}%
	\label{fig:SI:QO_DL}%
\end{figure}

Effective masses, \mstar\ have been determined from Lifshitz-Kosevich fits to the temperature dependence of the amplitude determined from Fourier transforms (cf. \autoref{fig:SI:QO_DL}). 
For the various orbits, the effective masses range from $\mstar=\qtyrange{2}{12}{\me}$ as listed in \autoref{tab:SI:QO}. The highest mass of $\mstar=\SI{12.2(2)}{\me}$ is detected for \Fd\ as shown in \autoref{fig:SI:Fd_Mass}
We analyse the effective mass for field windows demarcated by the LT transitions visible in the data (see \autoref{sec:SI:Exp} and \autoref{tab:SI:LT}). The field range over which the mass is obtained is indicated by grey horizontal bars in \autoref{fig:MvsH}(b). 

\section{Analysis of field-dependent quantum oscillation frequencies}
\label{sec:SI:QO_LT}
\begin{figure}%
	\includegraphics[width=\figurewidth]{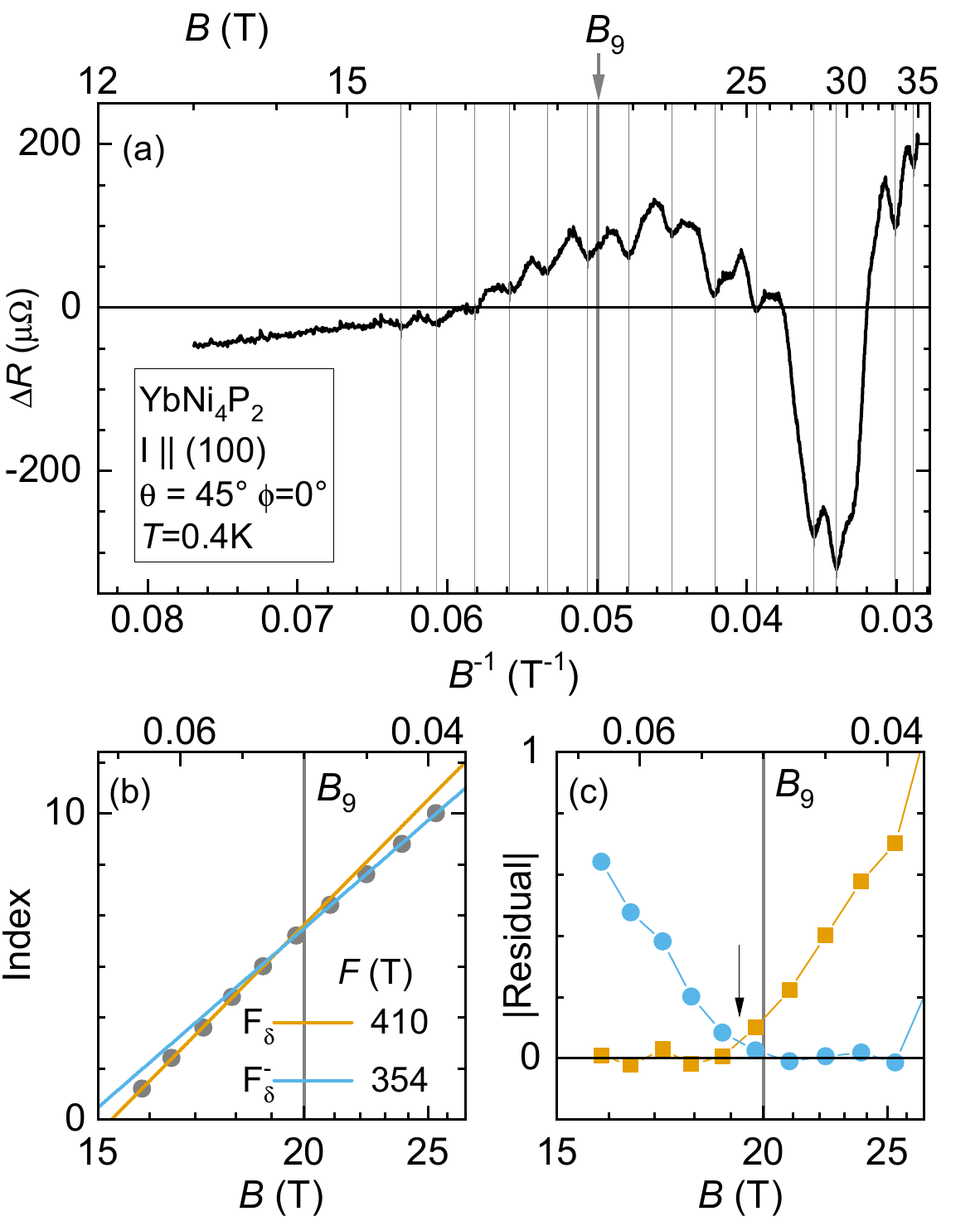}%
\caption{\textbf{Change of Quantum Oscillation Frequencies for large polar angle.} (a) Resistance after subtraction of linear background. Local minima in $R(B)$ are indicated by thin grey vertical lines. \Bix\ as determined previously  \cite{Karbassi2018} is  indicated as thick grey vertical line. (b) Indices of maxima from (a) vs inverse field together with linear fits for $B<\Bix$ (yellow line) and $B>\Bix$ (blue line). The slope of the lines in (b) corresponds to the quantum oscillation frequencies listed in the figure legend. (c) Residual of linear fits from (b). Arrow in (c) indicates the crossover associated with the change in quantum frequency. }%
\label{fig:SI:QO_LTc}%
\end{figure}

The indexing of the extrema in the quantum oscillations is used as a method to detect the change of quantum oscillation frequencies with minimal bias from the raw data. The maxima in \autoref{fig:QO_LTa}(a) and minima in \autoref{fig:SI:QO_LTc}(a) are directly visible in the data and are used to index the fields at which Landau levels pass through the Fermi surface at extremal orbits as confirmed by the scaling of the Landau levels with inverse magnetic field (\autoref{fig:QO_LTa}(b) and \autoref{fig:SI:QO_LTc}(b)). The gradient of the Landau-level index versus field corresponds to the quantum oscillation frequency of the dominant orbit. We observe a change in slope associated with \Bix\ as marked by the crossing of the linear fits to the low-field and high-field regimes in \autoref{fig:QO_LTa}(b) and \autoref{fig:SI:QO_LTc}(b). The crossing is in very good agreement with \Bix\ for low polar angles (\autoref{fig:QO_LTa}) and shifts clearly to higher fields for larger polar angles, (\autoref{fig:SI:QO_LTc}). For higher polar angles,  we note a deviation between the crossing and \Bix\ within the uncertainty of \Bix\ arising from the uncertainty in sample orientation and the steep evolution of $\Bix(\theta)$.

\section{Spin-Splitting model}
\label{sec:SI:SpinSplit}


In model B), we describe the non-linear spin splitting as illustrated in \autoref{fig:SpinSplit}(b). We assume a zero-field frequency \Fo, and a rate of change $a_1$ and $a_2$ due to Zeeman shift below and above \Bix. In addition, we consider a rate of change $c$ of the mean surface area reflecting the suppression of the Kondo effect. Continuous evolution of the areas requires for the Onsager frequencies
\begin{align}
	\FApm = 
		\begin{cases}
			\Fo \pm a_1 B - c B, B<\Bix \\
			\Fo\pm(a_1-a_2)\Bix \pm a_2 B , B>\Bix 
		\end{cases}
\label{eq:FA}
\end{align}
as shown by dotted lines in \autoref{fig:FSscenarios}(b).
The back-projected \Fbp\ frequencies corresponding to this model are given by 
\begin{align}
	\Fbppm = 
		\begin{cases}
			\Fo, B<\Bix \\
			\Fo \pm(a_1-a_2)\Bix - c\Bix  , B>\Bix 
		\end{cases}
\label{eq:g_Fbp}
\end{align}
and are shown in \autoref{fig:SpinSplit}(c). 
Thus, the difference between the back-projected frequencies is given by
\begin{equation}
	\Delta \Fbp = \Fbp^+ - \Fbp^- = 2\Bix(a_1-a_2)
\label{eq:DFbp}
\end{equation}
Assuming a simple free-electron model, we find that the difference in the product of effective masses and $g$-factor before and after \Bix\ is given by
\begin{equation}
	(\mstar_2 g_2 - \mstar_1 g_1) = \frac{\Delta\Fbp}{2\Bix}* \frac{\hbar e}{\me\muB m_s}
\label{eq:Dmg}
\end{equation}
where $\mstar_{(1,2)}$ and $g_{(1,2)}$ are the effective mass and $g$-factor below and above \Bix. $m_s = 5/2$ denotes dominant the spin state in the CEF. Using the effective masses observed in experiment, we find a change of the $g$ factor 
\begin{equation}
	\Delta g = g_2-g_1 = 0.5(2)
\label{eq:Dg}
\end{equation}

\section{Alternative explanations for spin-split frequencies}
\label{sec:SI:alternatives}

We can  rule out a Pomeranchuk instability as the cause for the splitting of the quantum oscillation frequencies. A Pomeranchuk instability arises, when a flat section of one of the spin branches passes through the chemical potential as a consequence of Zeeman shifts. This is expected to lead to a smooth variation of the quantum oscillation frequencies with a pronounced peak in the opposite direction of the final frequency shift as observed for instance in CeCoIn$_5$ \cite{Hornung2021}. We do not observe such a peak but rather our indexing analysis in \autoref{fig:QO_LTa} highlights the abrupt change of the quantum oscillation frequency over a small field range without a peak. It is also very unlikely, that we miss such a peak as the small field range of the frequency change would imply an unphysically sharp change of the electron dispersion relation. Finally, the anisotropy of the frequency splitting is hard to reconcile with a Pomeranchuk instability.

\section{Computational Details}
\label{sec:SI:DFT}
We approximate the band structure of \YNP\ by confining the $f$ electron to the core states. This is implemented by calculating the non-magnetic analogue \LNP\ using the experimental lattice parameters of \YNP\ \cite{Krellner2011}. For \LNP\ the $f$ electron shell is filled and situated well below the Fermi level. Wien2k was used to compute the band structure of \LNP\ \cite{Blaha2019}. The generalised gradient approximation of the exchange-correlation energies \cite{Perdew1996}. Valence states were treated scalar-relativistically, with spin–orbit coupling included in a second-variational step. The band structure was converged for a RKmax = 9 and 364 points in the irreducible wedge of the Brillouin zone (iBZ). A lower energy cut off of \SI{5}{Ry} resulted in an itinerant treatment of Lu 5$s$ 5$p$ 4$f$ 5$d$ 6$s$, Ni 3$d$ 4$s$, and P 3$s$ 3$p$ states. 
After convergence, band energies were evaluated on a finer grid of typically $\approx 7000$ points in the iBZ. From the band energies, the angular dependence of the quantum oscillation frequencies  were computed using SKEAF \cite{Rourke2012}. 

\begin{figure}
    \centering
    \includegraphics[width=\figurewidth]{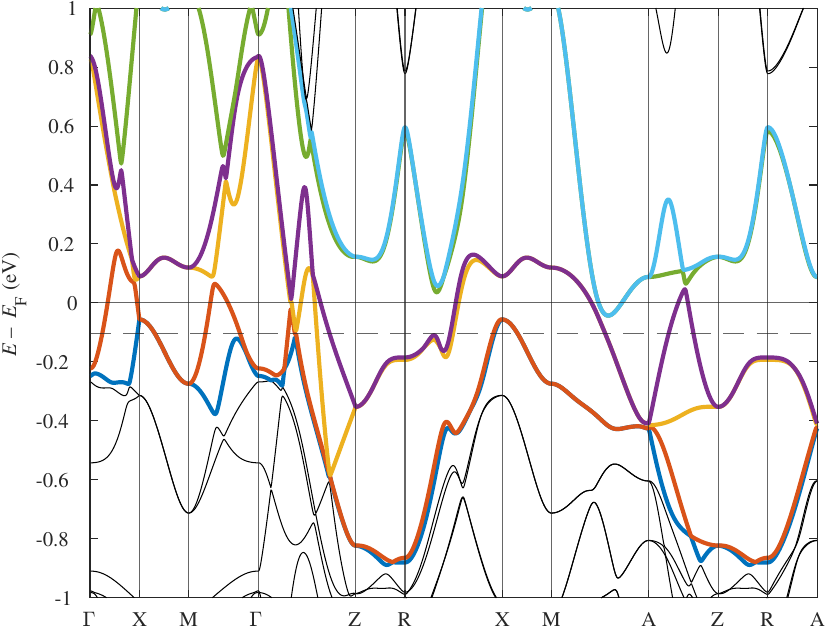}
    \caption{\textbf{Band structure} from spin-orbit DFT calculation of $f$-core approximation calculated using the electron configuration of \LNP\ and the experimental lattice constants of \YNP. A dotted line shows the shift equivalent to adding one hole.}
    \label{fig:SI:Spaghetti_LNP}
\end{figure}

We present the computed band structure in \autoref{fig:SI:Spaghetti_LNP}. This method provides the $f$-core Fermi surface equivalent to a fully localised Yb $f$ moment and hence does not include the $f$ degree of freedom in the Fermi surface volume. This situation is equivalent to a suppression of the Kondo effect that might take place at high magnetic fields \cite{Naren2013}. To account for a partial hybridisation of $f$ states with conduction electrons, we adjust the Fermi level in the evaluation of the Fermi surface topology and quantum oscillation frequencies (see \autoref{tab:SI:QO}). In the case of Yb-based heavy fermion materials, the Fermi energy is shifted to lower energies equivalent to a hole doping associated with the $f$-hole configuration of Yb \cite{Rourke2009}.

%

\begin{table}%
\begin{tabular}{l|c|l}
Band & $\Delta E$ & Colour \\ \hline
1    & \SI{-15}{\milli\electronvolt}      & Orange \\
2    & \SI{-20}{\milli\electronvolt}          & Yellow \\
3    & \SI{-35}{\milli\electronvolt}     & Purple \\
4    & \SI{-30}{\milli\electronvolt}     & green    \\
5    & \SI{-30}{\milli\electronvolt}      & blue
\end{tabular}
\caption{\textbf{Rigid band shifts} in DFT calculations and colour scheme used in \autoref{fig:FS} and \autoref{fig:rot_FS}.}
	\label{tab:SI:DFTShifts}
\end{table}%

\end{document}